\documentclass[referee]{gji}
\usepackage{flushend}
\usepackage{timet}
\usepackage{color}
\usepackage{balance}
\usepackage{amsmath}    
\usepackage{amssymb}    
\usepackage{mathrsfs}  
\usepackage{graphicx}
\usepackage{subfigure}
\usepackage{epstopdf}
\usepackage{ulem}

\newcommand{\ts}[1]{\boldsymbol {\mathsf #1}}

\newcommand{\co}[1]{\color{blue} #1}

\allowdisplaybreaks[4]
\title{Exact Closed-form Solutions for Lamb's Problem}
\author[Xi Feng and Haiming Zhang]{Xi Feng and Haiming Zhang \\
  Department of Geophysics, School of Earth and Space Sciences, Peking University \emph{100871}, P. R. China.\\
  Email: zhanghm@pku.edu.cn}
\date{In original form 2017 May 18}
\pagerange{\pageref{firstpage}--\pageref{lastpage}}
\volume{200}
\pubyear{2017}

\begin{document}

\label{firstpage}

\maketitle

\begin{summary}
In this article, we report on an exact closed-form solution for the displacement at the surface of an
elastic half-space elicited by a buried point source that acts at some point underneath that surface. This
is commonly referred to as the 3-D Lamb's problem, for which previous solutions were restricted to
sources and receivers placed at the free surface. By means of the reciprocity theorem, our solution
should also be valid as a means to obtain the displacements at interior points when the source is placed
at the free surface. We manage to obtain explicit results by expressing the solution in terms of
elementary algebraic expression as well as elliptic integrals. We anchor our developments on Poisson's
ratio 0.25 starting from Johnson's ({\co 1974}) integral solutions which must be computed numerically. In the end, our closed-form
results agree perfectly with the numerical results of Johnson ({\co 1974}), which strongly confirms the
correctness of our explicit formulas. It is hoped that in due time, these formulas may constitute a
valuable canonical solution that will serve as a yardstick against which other numerical solutions can be
compared and measured.
\end{summary}


\section{Introduction}
Beginning with Lamb's ({\color{blue}1904}) seminal work, Lamb's problem, which refer to dynamics response on the free surface of an elastic half-space resulting from a time-dependent point pulse embedded in the solid or just located on the free surface, has been the classic subject of numerous studies in theoretical seismology. Cagniard ({\color{blue}1939}) provided an intricate method that employs the Laplace tranform with respect to time and presented the final solutions in the time domain. 
Due to the modification by de Hoop ({\color{blue}1960}), it has become an appropriate way to solve Lamb's problem, which is now referred to as the ``Cagniard-de Hoop method". Garvin (1956) followed Cagniard's method to provide the exact displacements on the surface of a 2-D elastic half-space due to a suddenly applied, buried line of pressure. So the 2-D version of Lamb's problem is called Garvin's problem. Garvin's solution was extended by Alterman and Loewenthal (1969) to receivers in the interior, but their solution remained largely unnoticed. The problem was taken up again by S\'{a}nchez-Sesma \textit{et al} (2013), who provided a full set of formulas with the exact solution for arbitrary locations of the source and the receiver.
Kausel (2006) collected the classical solution for a horizontal or vertical line load applied in the interior, where the displacements are found on the free surface. The solution has been generalized by Tsai and Ma (1991), where the source and the receiver can be at any arbitrary location. But their work is obscure and they omitted the horizontal displacement component due to a horizontal load.
Ma and Huang (1996) also provided a 2-D solution for a line load acting within one of two adjoined half spaces of different material properties. Apparently, they also provided therein the component that was missing in Tsai and Ma (1991), but used a different coordinate system.

For the surface point pulse acting on a 3-D medium, Pekeris ({\color{blue}1955a}) gave a closed-form solution for a vertical point source and a tangential point force on the surface was treated by Chao ({\color{blue}1960}). They provided exact results when Poisson's ratio is 0.25, and Mooney ({\color{blue}1974}) extended the results for vertical loads with an arbitrary Poisson's ratio, while ignoring the radial component. Starting with Johnson's ({\color{blue}1974}) results, Richards ({\color{blue}1979}) obtained the complete set of formulae for studying spontaneous crack propagation, but providing no proofs or derivations. Recently, Kausel ({\color{blue}2012}) obtained these formulae again with more details and his formulae are much simpler than Richards's ({\color{blue}1979}), besides, exact formulas for six of the nine possible dipoles that can act on the surface of the medium were also provided. The special case of Lamb's problem for the displacements on the surface elicited by a surface point load has been solved completely by Kausel ({\color{blue}2012}).

For a point pulse at depth, Pinney ({\color{blue}1954}), Pekeris ({\color{blue}1955b}), Pekeris and Lifson ({\color{blue}1957}) and  Aggarwal and Ablow ({\color{blue}1967}) obtained a class of results in terms of finite integrals. Pekeris ({\color{blue}1955b}) and Pekeris and Lifson ({\color{blue}1957}) gave exact integral solutions of the displacements produced by a point pressure pulse and a vertical force, respectively. An asymptotic formula for the Rayleigh wave effects was given in Aggarwal and Ablow ({\color{blue}1967}). 
Kawasaki \textit{et al} ({\color{blue}1972a, 1972b}) and Sato ({\color{blue}1972}) solved the surface displacements due to a double-couple source and a fault model. Johnson ({\color{blue}1974}) collected these solutions together with a uniform notation and in a form suitable for numerical calculations. He presented the Green's function and its spatial derivatives in terms of finite integrals by the Cagniard-de Hoop method. The solutions for the more general situation where the receiver points are also located at depth, was obtained in his article, though in cumbersome forms. Roy ({\color{blue}1974}) considered a moving point source with constant velocity along a line that is inclined at some angle with respect to the horizontal. He allegedly gave the displacements on the surface in ``closed-form'', but actually his formulas were expressed in terms of integrals that must be evaluated numerically.

Due to the integral form of these results, one has to draw support from asymptote methods (Lapwood, {\color{blue}1949}) and numerical computations (Pekeris and Lifson, {\color{blue}1957}), in order to determine the detailed properties of the waves. So far a few studies give the properties. In this paper, we decompose the integral solutions by Johnson ({\color{blue}1974}) into elementary algebraic expressions and standard elliptic integrals. An algebraic expression relates to a mathematical expression or equation in which a finite number of symbols is combined using only the operations of addition, subtraction, multiplication, division, and exponentiation with constant rational exponents. Simple and exact expressions for the Rayleigh waves and the diffracted S-P wave, which was first pointed out in Nakano ({\color{blue}1925}) for the 2-D case, are finally obtained. Our work enables us to analyze the basic features of these waves with precise calculations and a rigorous mathematical structure.

It should be pointed out that although we focus on the case with a Poisson's ratio 0.25, which is an adequate approximate value for the Earth (Lapwood, {\color{blue}1949}), our formulae are valid for a large range of Poisson's ratios ($0<\nu<0.2631$). It should be noted that the Rayleigh function has two conjugate complex roots for other Poisson's ratios, and hence the final expressions are somewhat cumbersome.
\section{Notation and definitions}
In this paper, we consider the solution of the 3-D problem, with the displacement at the free surface in an isotropic elastic solid with the Lam\'{e} constants $\lambda$ and $\mu$ and a density $\rho$. Since the simplest source, the unidirectional unit impulse, which is localized precisely in both space and time is considered, the displacement field from such a simple source is the elastodynamic Green's function. A Cartesian coordinate system is established such that the free surface is located in the plane $x_3=0$, and the positive $x_3$ axis points downward (Fig. 1).
The source is fixed on the $x_3$ axis, and the receiver is on the free surface. We denote as $r$ the source-receiver distance, and $\theta$ is the angle between the vertical $x-z$ plane containing the source and the vertical plane containing the receiver. We refer to this angle as the azimuth, even if herein it is defined to be positive in the clockwise direction.\\
The following symbols are used in the ensuing:\\
\begin{tabular}{|c|l|}
\hline
$r$ & distance between source and receiver \\
$\theta$ & azimuth\\
$\phi$ & polar angle\\
$\lambda$,\,$\mu$ & Lam\'{e} constants \\
$\rho$ & mass density \\
$t$ & time \\
$\alpha$ & velocity of P wave \\
$\beta$ & velocity of S wave \\
$k=\frac{\alpha}{\beta}$ & ratio of P to S wave velocity \\
$\ts{G}$ & A tensor with a name G\\
$T_{\rm P}=\frac{\alpha t}{r}$ & dimensionless time in $\ts{G}^{\rm P}$ \\
$T_{\rm S}=\frac{\beta t}{r}$ & dimensionless time in $\ts{G}^{\rm S}$ and $\ts{G}^{\rm S\text{-}P}$\\
($x'_{1}$, $x'_{2}$, $x'_{3}$) & coordinate of the source\\
($x_{1}$, $x_{2}$, $x_{3}$) & coordinate of the receiver\\
\hline
\end{tabular}
\begin{figure}
\centering
\includegraphics[width=.6\textwidth]{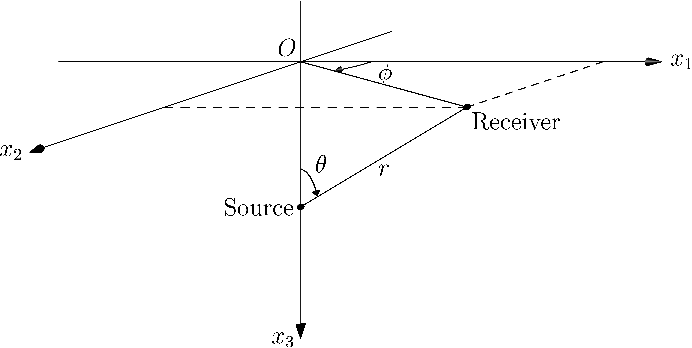}
\caption{The geometry of the problem. The Cartesian coordinate used in this paper. The coordinate of the source is ($0$, $0$, $x'_{3}$) and the coordinate of the receiver is ($x_{1}$, $x_{2}$, 0) since we only
discuss the case when the receiver is on the free surface. The distance between the source and the receiver is $r$.}
\end{figure}
\section{Introduction to Johnson's result}
Johnson ({\color{blue}1974}) obtained the representation of the 3-D Green's function in a uniform elastic half-space.
Since his original expressions are not convenient for our method, we rewrite his formulae as follows:
\begin{align}
\ts{G}=\ts{G}^{\rm P}+\ts{G}^{\rm S}+\ts{G}^{\rm S\text{-}P}
\end{align}
\begin{align}
\label{eqn1}\ts{G}^{\rm P}=\frac{1}{\pi^{2}\mu r}&\frac{\partial}{\partial t}\left.\int^{p_{_{\rm P}}(t)}_{0}H\left(t-t_{_{\rm P}}\right)\Re\left[ \frac{\eta_\alpha\ts{P}(p,q)}{\sigma\sqrt{p^2_{_{\rm P}}(t)-p^2}}\right]\right|_{q=q_{_{\rm P}}(t,p)}\mathrm{d}p, \\
\label{eqn2}\ts{G}^{\rm S}=\frac{1}{\pi^{2}\mu r}&\frac{\partial}{\partial t}\left\{\left.\int^{p_{_{\rm S}}(t)}_{0}H\left(t-t_{_{\rm S}}\right)\Re\left[ \frac{\eta_{_\beta}\ts{S}(p,q)}{\sigma\sqrt{p^2_{_{\rm S}}(t)-p^2}}\right]\right|_{q=q_{_{\rm S}}(t,p)}\mathrm{d}p \right. \notag \\
&\left.-H\left(\sin\theta-\tfrac{1}{k}\right)\left.\int_{p_{_{\rm S}}(t)}^{p_{_{\rm S\text{-}P}}(t)}H\left(t-t_{_{\rm S}}\right)\Im\left[\frac{\eta_{_\beta}\ts{S}(p,q)}{\sigma\sqrt{p^2-p^2_{_{\rm S}}(t)}}\right]\right|_{q=q_{_{\rm S\text{-}P}}(t,p)}\mathrm{d}p\right\},  \\
\label{eqn3}\ts{G}^{\rm S\text{-}P}=-\frac{1}{\pi^{2}\mu r}&\left.H\left(\sin\theta-\tfrac{1}{k}\right)\frac{\partial}{\partial t}\int^{p_{_{\rm S\text{-}P}}(t)}_{0}\left[H\left(t-t_{_{\rm S\text{-}P}}\right)-H\left(t-t_{_{\rm S}}\right)\right]\Im\left[\frac{\eta_{_\beta}\ts{S}(p,q)}{\sigma\sqrt{p^2-p^2_{_{\rm S}}(t)}}\right]\right|_{q=q_{_{\rm S\text{-}P}}(t,p)}\mathrm{d}p,
\end{align}
where $H(\cdot)$ is the Heaviside step function, $\Re\{\cdot\}$ and $\Im\{\cdot\}$ represent the real and imaginary part, respectively,
\begin{align*}
&t_{_{\rm P}}=\frac{r}{\alpha}, && t_{_{\rm S}} =\frac{r}{\beta}, && t_{_{\rm S\text{-}P}}=\frac{r}{\alpha}\sin\theta+r\sqrt{\frac{1}{\beta^2}-\frac{1}{\alpha^2}}\cos\theta, \notag \\
&p_{_{\rm P}}(t)=\sqrt{\frac{t^2}{r^2}-\frac{1}{\alpha^2}},&& p_{_{\rm S}}(t)=\sqrt{\frac{t^2}{r^2}-\frac{1}{\beta^2}}, &&
p_{_{\rm S\text{-}P}}(t)=\sqrt{\left(\frac{\frac{t}{r}-\sqrt{\frac{1}{\beta^2}-\frac{1}{\alpha^2}}\cos\theta}{\sin\theta}\right)^2
-\frac{1}{\alpha^2}}, \notag \\
&\eta_\alpha=\sqrt{\frac{1}{\alpha^2}+p^2-q^2}, &&\eta_{_\beta}=\sqrt{\frac{1}{\beta^2}+p^2-q^2},
&&\sigma=\left(\eta_{_\beta}^2+p^2-q^2\right)^2+4\eta_\alpha\eta_{_\beta}\left(q^2-p^2\right),
\end{align*}
and
\begin{align*}
q_{_{\rm P}}(t,p)&=-\frac{t}{r}\sin{\theta}+{\rm i}\sqrt{p^2_{_{\rm P}}(t)-p^2}\cos\theta, \notag \\
q_{_{\rm S}}(t,p)&=-\frac{t}{r}\sin{\theta}+{\rm i}\sqrt{p^2_{_{\rm S}}(t)-p^2}\cos\theta, \notag \\
q_{_{\rm S\text{-}P}}(t,p)&=-\frac{t}{r}\sin{\theta}+\sqrt{p^2-p^2_{_{\rm S}}(t)}\cos\theta.
\end{align*}
The expressions for the individual elements of the three-by-three matrices $\ts{P}$ and $\ts{S}$ are as follows:
\begin{align}
\ts{P}&=\begin{bmatrix}
 2\eta_{_\beta}\epsilon & 2\eta_{_\beta}\zeta & 2q\eta_\alpha\eta_{_\beta}\cos\phi \\
 2\eta_{_\beta}\zeta & 2\eta_{_\beta}\bar{\epsilon} & 2q\eta_\alpha\eta_{_\beta}\sin\phi \\
 q\gamma\cos\phi & q\gamma\sin\phi & \gamma\eta_\alpha
\end{bmatrix}, \\
\ts{S}&=\frac{1}{\eta_{_\beta}}\begin{bmatrix}
 \eta_{_\beta}^2\gamma-\bar{\gamma}\bar{\epsilon} & \bar{\gamma}\zeta & -q\eta_{_\beta}\gamma\cos\phi \\
 \bar{\gamma}\zeta & \eta_{_\beta}^2\gamma-\bar{\gamma}\epsilon & -q\eta_{_\beta}\gamma\sin\phi \\
 -2q\eta_\alpha\eta^2_{_\beta}\cos\phi & -2q\eta_\alpha\eta^2_{_\beta}\sin\phi & 2\eta_\alpha\eta_{_\beta}\left(q^2-p^2\right)
\end{bmatrix}, 
\end{align}
where
\begin{align*}
 \gamma&=\eta_{_\beta}^2+p^2-q^2,  &\bar{\gamma}&=\gamma-4\eta_\alpha\eta_{_\beta},  \notag \\
 \epsilon&=q^2\cos^2\phi-p^2\sin^2\phi, &\bar{\epsilon}&=q^2\sin^2\phi-p^2\cos^2\phi, \notag \\
 \zeta&=\left(q^2+p^2\right)\sin\phi\cos\phi.
\end{align*}
 $\sigma$ could be defined as the Rayleigh function (Aki and Richards, {\color{blue}2002}). Let $y=\beta^2(q^2-p^2)$ (Liu \textit{et al},  
 {\color{blue}2016}), then 
\begin{align}
\sigma(y)=\frac{(1-2y)^2+4y\sqrt{k^{-2}-y}\sqrt{1-y}}{\beta^4}. 
\end{align}
Let 
\begin{align*}
&\sigma_{1}=(1-2y)^2,\\
&\sigma_{2}=4y\sqrt{k^{-2}-y}\sqrt{1-y},
\end{align*}
then the zero of $\sigma$ must be the zero of $\sigma_{1}^2-\sigma_{2}^2$, which is a cubic polynomial of $y$
\begin{align}
\label{ray}R(y)=-16(1-k^{-2})y^3+8(3-2k^{-2})y^2-8y+1=0.
\end{align}
The nature of the three roots to the cubic equation ({\ref{ray}}) depend on the so-called discriminant of the cubic equation, the value of which depends in turn on the parameter $k$. If $1.4142<k<1.765$ (i.e. $0<\nu<0.2631$), all three roots are real, otherwise there is one real root and two complex conjugate roots.
In the second situation the latter calculation is elaborate, though our method is still valid. Due to limited space, explicit discussions of this situation will be omitted in this article. We concentrate on the first situation that contains the case with a Poission's ratio 0.25, which is a proper approximate value for the Earth. Then we analyze the distributions of the real roots. Note that $R(0)=R(1)=1$ and $R(k^{-2})=(2k^{-2}-1)^4\geq 0$, the euqation has two roots $y_{1}$ and $y_{2}$ both falling in $[0, k^{-2}]$ and the other root $y_{3}$ greater than $1$. $y_{3}$ is the actual zero of the Rayleigh function $\sigma(y)$ because $\sigma(y_{1})$ and $\sigma(y_{2})$ are obviously positive. So $y_{3}$ associates with the generation of the Rayleigh wave.
 
In the following sections, we will tranform $\ts{G}^{\rm P}$, $\ts{G}^{\rm S}$ and $\ts{G}^{\rm S\text{-}P}$ into expressions with elementary algebraic functions and elliptic integrals.
\section{Calculation of $\ts{G}^{\rm P}$}
In this section, we decompose $\ts{G}^{\rm P}$ into elementary algebraic expressions and standard elliptic integrals.
The whole process is complicated so we should list the essential steps of the process: 
\begin{enumerate}
\item Introduce a new variable $B=\alpha\eta_{\alpha}$ as the integral variable. With the substitution, we convert the form of the original expressions into two terms: the first term can be integrated in terms of elementary algebraic functions and the second can be expressed in terms of standard elliptic integrals. 
\item Decompose each term as a linear combination of several integrations abbreviated as $U_i^{\rm P}\ (i=1,2,\cdots,6)$ and $V_i^{\rm P}\ (i=1,2,\cdots,7)$ by applying the polynomial division. The corresponding coefficients are easily obtained using the symbolic tool in Matlab. 
\item Calculate every $U_i^{\rm P}$ and $V_i^{\rm P}$. We clearly outline in the ensuing the procedure while omitting the detailed derivations.
\end{enumerate} 
In order to transform equation ({\ref{eqn1}}) into a form easy to integrate, we introduce a new variable $B=\alpha\eta_{\alpha}$, hence
\begin{align*}
&p=\sqrt{B^{2}-2T_{\rm P}\cos{\theta}B+T_{\rm P}^{2}-\sin^{2}{\theta}}/(\alpha\sin{\theta}), && q=\frac{\cos{\theta}B-T_{\rm P}}{\alpha\sin{\theta}},\\
&\eta_{_\beta}=\sqrt{B^{2}+k^{2}-1}/\alpha, && \gamma=(2B^{2}+k^{2}-2)/\alpha^{2},\\
&\frac{1}{\sigma}=\frac{(2B^{2}+k^{2}-2)^{2}+4B(B^{2}-1)\sqrt{B^{2}+k^{2}-1}}{(2B^{2}+k^{2}-2)^{4}-16B^{2}(B^{2}+k^{2}-1)(B^{2}-1)^{2}}\alpha^{4}, && p^{2}-q^{2}=(B^{2}-1)/\alpha^{2},
\end{align*}
\begin{align*}
\int^{p_{_{\rm P}}(t)}_{0}\frac{\mathrm{d}p}{\sqrt{p_{_{\rm P}}^{2}(t)-p^{2}}}=\int_{T_{\rm P}\cos{\theta}}^{T_{\rm P}\cos{\theta}+{\rm i}\sqrt{T_{\rm P}^{2}-1}\sin{\theta}}\frac{-{\rm i}\mathrm{d}B}{\sqrt{B^{2}-2T_{\rm P}\cos{\theta}B+T_{\rm P}^{2}-\sin^{2}{\theta}}}.
\end{align*}
By these substitutions, equation ({\ref{eqn1}}) can be transformed into
\begin{small}
\begin{align}
\ts{G}^{\rm P}=&\frac{1}{\pi^{2}\mu r}\frac{\partial}{\partial t}\left(\Im\int_{T_{\rm P}\cos{\theta}}^{T^{\rm P}_{\rm upper}}H\left(t-t_{_{\rm P}}\right)\frac{\ts{M}^{\rm P}(B)}{R^{\rm P}\left(B\right)}\frac{\mathrm{d}B}{\sqrt{Q^{\rm P}_{1}\left(B\right)}}+\Im\int_{T_{\rm P}\cos{\theta}}^{T^{\rm P}_{\rm upper}}H\left(t-t_{_{\rm P}}\right)\frac{\ts{N}^{\rm P}\left(B\right)}{R^{\rm P}\left(B\right)}\frac{\mathrm{d}B}{\sqrt{Q^{\rm P}_{2}\left(B\right)}}\right),
\end{align}
\end{small}
where 
\begin{small}
\begin{align}
\label{RP}&T^{\rm P}_{\rm upper}=T_{\rm P}\cos{\theta}+{\rm i}\sqrt{T_{\rm P}^{2}-1}\sin{\theta}, &&
R^{\rm P}(B)=(2B^{2}+k^{2}-2)^{4}-16B^{2}(B^{2}+k^{2}-1)(B^{2}-1)^{2},\\
\label{QP}&Q^{\rm P}_{1}(B)=B^{2}-2T_{\rm P}\cos{\theta}B+T_{\rm P}^{2}-\sin^{2}{\theta}, && 
Q^{\rm P}_{2}(B)=(B^{2}+k^{2}-1)(B^{2}-2T_{\rm P}\cos{\theta}B+T_{\rm P}^{2}-\sin^{2}{\theta}).
\end{align}
\end{small}
As a cubic polynomial of $B^2$, $R^{\rm P}(B^2)$ is equivalent to the Rayleigh function $\sigma(y)$ noticing that $y=k^{-2}(1-B^2)$. So $R^{\rm P}(B^2)$ has a negative root $-(a^{\rm P}_{3})^2=1-k^2y_{3}$ (actual zero) and two positive roots $(a^{\rm P}_{1})^2=1-k^2y_{1}$ and $(a^{\rm P}_{2})^2=1-k^2y_{2}$. $a^{\rm P}_{1}, a^{\rm P}_{2},a^{\rm P}_{3}$ are defined such that they are all positive numbers.

All the individual elements of the matrices $\ts{M}^{\rm P}(B)$ are 8-th order polynomials of $B$, and those of the matrices $\ts{N}^{\rm P}(B)$ are 9-th order polynomials of $B$. Detailed expressions of $\ts{M}^{\rm P}(B)$ and $\ts{N}^{\rm P}(B)$ are listed as follows,
\begin{align}
\label{Mfirst}M^{\rm P}_{11}&=I^{\rm P}_{1}\cos^{2}{\phi}-I^{\rm P}_{2}\sin^{2}{\phi},\\
M^{\rm P}_{22}&=I^{\rm P}_{1}\sin^{2}{\phi}-I^{\rm P}_{2}\cos^{2}{\phi},\\
M^{\rm P}_{12}&=M^{\rm P}_{21}=(I^{\rm P}_{1}+I^{\rm P}_{2})\sin{\phi}\cos{\phi},\\
M^{\rm P}_{13}&=\frac{8}{\sin{\theta}}B^{3}(B^{2}+k^{2}-1)(B^{2}-1)(\cos{\theta}B-T_{\rm P})\cos{\phi},\\
M^{\rm P}_{23}&=\frac{8}{\sin{\theta}}B^{3}(B^{2}+k^{2}-1)(B^{2}-1)(\cos{\theta}B-T_{\rm P})\sin{\phi},\\
M^{\rm P}_{31}&=\frac{1}{\sin{\theta}}B(2B^{2}+k^{2}-2)^{3}(\cos{\theta}B-T_{\rm P})\cos{\phi},\\
M^{\rm P}_{32}&=\frac{1}{\sin{\theta}}B(2B^{2}+k^{2}-2)^{3}(\cos{\theta}B-T_{\rm P})\sin{\phi},\\
\label{Mlast}M^{\rm P}_{33}&=B^{2}(2B^{2}+k^{2}-2)^{3},
\end{align}
where
\begin{align*}
I^{\rm P}_{1}&=\frac{8}{\sin^{2}{\theta}}B^{2}(B^{2}+k^{2}-1)(B^{2}-1)(\cos{\theta}B-T_{\rm P})^{2},\\
I^{\rm P}_{2}&=\frac{8}{\sin^{2}{\theta}}B^{2}(B^{2}+k^{2}-1)(B^{2}-1)(B^{2}-2T_{\rm P}\cos{\theta}B+T_{\rm P}^{2}-\sin^{2}{\theta}),
\end{align*}
and
\begin{align}
\label{Nfirst}N^{\rm P}_{11}&=I^{\rm P}_{3}\cos^{2}{\phi}-I^{\rm P}_{4}\sin^{2}{\phi},\\
N^{\rm P}_{22}&=I^{\rm P}_{3}\sin^{2}{\phi}-I^{\rm P}_{\rm 4}\cos^{2}{\phi},\\
N^{\rm P}_{12}&=N^{\rm P}_{21}=(I^{\rm P}_{3}+I^{\rm P}_{4})\sin{\phi}\cos{\phi},\\
N^{\rm P}_{13}&=(2B^{2}+k^{2}-2)I^{\rm P}_{5}\cos{\phi},\\
N^{\rm P}_{23}&=(2B^{2}+k^{2}-2)I^{\rm P}_{5}\sin{\phi},\\
N^{\rm P}_{31}&=2(B^{2}-1)I^{\rm P}_{5}\cos{\phi},\\
N^{\rm P}_{32}&=2(B^{2}-1)I^{\rm P}_{5}\sin{\phi},\\
\label{Nlast}N^{\rm P}_{33}&=4B^{3}(B^{2}+k^{2}-1)(2B^{2}+k^{2}-2)(B^{2}-1),
\end{align}
where
\begin{align*}
I^{\rm P}_{3}&=\frac{2}{\sin^{2}{\theta}}B(2B^{2}+k^{2}-2)^{2}(B^{2}+k^{2}-1)(\cos{\theta}B-T_{\rm P})^{2},\\
I^{\rm P}_{4}&=\frac{2}{\sin^{2}{\theta}}B(2B^{2}+k^{2}-2)^{2}(B^{2}+k^{2}-1)(B^{2}-2T_{\rm P}\cos{\theta}B+T_{\rm P}^{2}-\sin^{2}{\theta}),\\
I^{\rm P}_{5}&=\frac{2}{\sin{\theta}}B^{2}(B^{2}+k^{2}-1)(2B^{2}+k^{2}-2)(\cos{\theta}B-T_{\rm P}).
\end{align*} 
Since $Q^{\rm P}_{1}(B)$ is a quadratic polynomial, the first term of $\ts{G}^{\rm P}$ can be integrated in terms of elementary functions. According to the algebraic function theory, the other term should be expressed by the elliptic integrals since $Q^{\rm P}_{2}(B)$ is a quartic polynomial. In the rest of this section, we proceed to implement these goals. First, we decompose every components of $\frac{M^{\rm P}_{ij}(B)}{R^{\rm P}(B)}$ and $\frac{N^{\rm P}_{ij}(B)}{R^{\rm P}(B)}$ into a series of monomials and partial fractions as 
\begin{align*}
\frac{M^{\rm P}_{ij}(B)}{R^{\rm P}(B)}=&\frac{u^{\rm P}_{ij,1}}{B-a^{\rm P}_{1}}+
\frac{u^{\rm P}_{ij,2}}{B+a^{\rm P}_{1}}+\frac{u^{\rm P}_{ij,3}}{B-a^{\rm P}_{2}}+\frac{u^{\rm P}_{ij,4}}{B+a^{\rm P}_{2}}+\frac{u^{\rm P}_{ij,5}}{B^{2}+\left(a^{\rm P}_{3}\right)^{2}}+\frac{u^{\rm P}_{ij,6}B}{B^{2}+\left(a^{\rm P}_{3}\right)^{2}} \nonumber \\
&+u^{\rm P}_{ij,7}+u^{\rm P}_{ij,8}B+u^{\rm P}_{ij,9}B^{2},\\
\frac{N^{\rm P}_{ij}(B)}{R^{\rm P}(B)}=&\frac{v^{\rm P}_{ij,1}}{B-a^{\rm P}_{1}}+
\frac{v^{\rm P}_{ij,2}}{B+a^{\rm P}_{1}}+\frac{v^{\rm P}_{ij,3}}{B-a^{\rm P}_{2}}+\frac{v^{\rm P}_{ij,4}}{B+a^{\rm P}_{2}}+\frac{v^{\rm P}_{ij,5}}{B^{2}+\left(a^{\rm P}_{3}\right)^{2}}+\frac{v^{\rm P}_{ij,6}B}{B^{2}+\left(a^{\rm P}_{3}\right)^{2}}\nonumber \\
&+v^{\rm P}_{ij,7}+v^{\rm P}_{ij,8}B+v^{\rm P}_{ij,9}B^{2}+v^{\rm P}_{ij,10}B^{3}.
\end{align*}
The coefficients $u_{ij,k}^{\rm P}$ and $v_{ij,k}^{\rm P}$ can be easily calculated by the polynomial division and the residue theorem of polynomials. The explicit expressions will be presented in Appendix A. Alternatively, all the coefficients can be evaluated directly by applying the Matlab command ``residue". 
Then the integration of $\ts{G}^{\rm P}$ can be seperated as the linear combination of $U_i^{\rm P}\ (i=1,2,\cdots,6)$ and $V_i^{\rm P}\ (i=1,2,\cdots,7)$
\begin{align}
\label{GP}G^{\rm P}_{ij}=&\frac{1}{\pi^{2}\mu r}\frac{\partial}{\partial t}\Big\{H\left(t-t_{_{\rm P}}\right)\Big[u^{\rm P}_{ij,1}U^{\rm P}_{1}\left(a^{\rm P}_{1}\right)+u^{\rm P}_{ij,2}U^{\rm P}_{1}\left(-a^{\rm P}_{1}\right)+u^{\rm P}_{ij,3}U^{\rm P}_{1}\left(a^{\rm P}_{2}\right)+u^{\rm P}_{ij,4}U^{\rm P}_{1}\left(-a^{\rm P}_{2}\right)\nonumber
\\&+u^{\rm P}_{ij,5}U^{\rm P}_{2}\left(a^{\rm P}_{3}\right)+u^{\rm P}_{ij,6}U^{\rm P}_{3}\left(a^{\rm P}_{3}\right)+u^{\rm P}_{ij,7}U^{\rm P}_{4}+u^{\rm P}_{ij,8}U^{\rm P}_{5}+u^{\rm P}_{ij,9}U^{\rm P}_{6}\Big]\Big\}\nonumber\\
&+\frac{1}{\pi^{2}\mu r}\frac{\partial}{\partial t}\Big\{H\left(t-t_{_{\rm P}}\right)\Big[v^{\rm P}_{ij,1}V^{\rm P}_{1}\left(a^{\rm P}_{1}\right)+v^{\rm P}_{ij,2}V^{\rm P}_{1}\left(-a^{\rm P}_{1}\right)+v^{\rm P}_{ij,3}V^{\rm P}_{1}\left(a^{\rm P}_{2}\right)+v^{\rm P}_{ij,4}V^{\rm P}_{1}\left(-a^{\rm P}_{2}\right)\nonumber
\\&+v^{\rm P}_{ij,5}V^{\rm P}_{2}\left(a^{\rm P}_{3}\right)+v^{\rm P}_{ij,6}V^{\rm P}_{3}\left(a^{\rm P}_{3}\right)+v^{\rm P}_{ij,7}V^{\rm P}_{4}+v^{\rm P}_{ij,8}V^{\rm P}_{5}+v^{\rm P}_{ij,9}V^{\rm P}_{6}+v^{\rm P}_{ij,10}V^{\rm P}_{7}\Big]\Big\},
\end{align}
where
\begin{align*}
&U^{\rm P}_{1}\left(a\right)=\Im\int_{T_{\rm P}\cos{\theta}}^{T_{\rm upper}^{\rm P}}\frac{1}{B-a}\frac{\mathrm{d}B}{\sqrt{Q^{\rm P}_{1}(B)}},
&& U^{\rm P}_{2}\left(a\right)=\Im\int_{T_{\rm P}\cos{\theta}}^{T_{\rm upper}^{\rm P}}\frac{1}{B^{2}+a^{2}}\frac{\mathrm{d}B}{\sqrt{Q^{\rm P}_{1}(B)}},\\
&U^{\rm P}_{3}\left(a\right)=\Im\int_{T_{\rm P}\cos{\theta}}^{T_{\rm upper}^{\rm P}}\frac{B}{B^{2}+a^{2}}\frac{\mathrm{d}B}{\sqrt{Q^{\rm P}_{1}(B)}},
&& U^{\rm P}_{4}=\Im\int_{T_{\rm P}\cos{\theta}}^{T_{\rm upper}^{\rm P}}\frac{\mathrm{d}B}{\sqrt{Q^{\rm P}_{1}(B)}},\\
&U^{\rm P}_{5}=\Im\int_{T_{\rm P}\cos{\theta}}^{T_{\rm upper}^{\rm P}}\frac{B\mathrm{d}B}{\sqrt{Q^{\rm P}_{1}(B)}},
&& U^{\rm P}_{6}=\Im\int_{T_{\rm P}\cos{\theta}}^{T_{\rm upper}^{\rm P}}\frac{B^{2}\mathrm{d}B}{\sqrt{Q^{\rm P}_{1}(B)}},\\
\end{align*}
and
\begin{align*}
&V^{\rm P}_{1}\left(a\right)=\Im\int_{T_{\rm P}\cos{\theta}}^{T_{\rm upper}^{\rm P}}\frac{1}{B-a}\frac{\mathrm{d}B}{\sqrt{Q^{\rm P}_{2}(B)}},
&& V^{\rm P}_{2}\left(a\right)=\Im\int_{T_{\rm P}\cos{\theta}}^{T_{\rm upper}^{\rm P}}\frac{1}{B^{2}+a^{2}}\frac{\mathrm{d}B}{\sqrt{Q^{\rm P}_{2}(B)}},\\
&V^{\rm P}_{3}\left(a\right)=\Im\int_{T_{\rm P}\cos{\theta}}^{T_{\rm upper}^{\rm P}}\frac{B}{B^{2}+a^{2}}\frac{\mathrm{d}B}{\sqrt{Q^{\rm P}_{2}(B)}},
&& V^{\rm P}_{4}=\Im\int_{T_{\rm P}\cos{\theta}}^{T_{\rm upper}^{\rm P}}\frac{\mathrm{d}B}{\sqrt{Q^{\rm P}_{2}(B)}},\\
&V^{\rm P}_{5}=\Im\int_{T_{\rm P}\cos{\theta}}^{T_{\rm upper}^{\rm P}}\frac{B\mathrm{d}B}{\sqrt{Q^{\rm P}_{2}(B)}},
&& V^{\rm P}_{6}=\Im\int_{T_{\rm P}\cos{\theta}}^{T_{\rm upper}^{\rm P}}\frac{B^{2}\mathrm{d}B}{\sqrt{Q^{\rm P}_{2}(B)}},\\
&V^{\rm P}_{7}=\Im\int_{T_{\rm P}\cos{\theta}}^{T_{\rm upper}^{\rm P}}\frac{B^{3}\mathrm{d}B}{\sqrt{Q^{\rm P}_{2}(B)}}.
\end{align*}
Next step we focus on the calculations of every $U_i^{\rm P}$ and $V_i^{\rm P}$. Our goal is to express all $U_i^{\rm P}\ (i=1,2,\cdots,6)$ in terms of elementary algebraic expressions and $V_i^{\rm P}\ (i=1,2,\cdots,6)$ in terms of elementary algebraic expressions as well as standard elliptic integrals.
\subsection{Calculation of $U^{\rm P}$}
With the substitution $B=T_{\rm P}\cos{\theta}+{\rm i}\sqrt{T_{\rm P}^{2}-1}\sin{\theta}\cos{x}$, the integrals $U_{i}^{\rm P}\ (i=1,2,\cdots,6)$ can be transformed into trigonometric integrals as
\begin{align*}
&U^{\rm P}_{1}\left(a\right)
=\Re\int_{0}^{\pi/2}\frac{\mathrm{d}x}{B-a},
&&U^{\rm P}_{2}\left(a\right)
=\Re\int_{0}^{\pi/2}\frac{\mathrm{d}x}{B^{2}+a^{2}},\\
&U^{\rm P}_{3}\left(a\right)
=\Re\int_{0}^{\pi/2}\frac{B\mathrm{d}x}{B^{2}+a^{2}},
&& U^{\rm P}_{4}
=\Re\int_{0}^{\pi/2}1 \mathrm{d}x,\\
&U^{\rm P}_{5}
=\Re\int_{0}^{\pi/2}B \mathrm{d}x,
&& U^{\rm P}_{6}
=\Re\int_{0}^{\pi/2}B^{2} \mathrm{d}x.
\end{align*}

By applying Mooney's ({\color{blue}1974}) method, we succeed to solve the integral $U_{1}^{\rm P}(a)$. The calculations of $U_{2}^{\rm P}(a)$ and $U_{3}^{\rm P}(a)$ are similar, and detailed deductions are demonstrated in Appendix B. The integrals $U_{4}^{\rm P}$, $U_{5}^{\rm P}$ and $U_{6}^{\rm P}$ can be easily obtained.

We enumerate the ultimate results of $U_i^{\rm }\ (i=1,2,\cdots,6)$ as follows:
\begin{align}
\label{Ufirst}&U_{1}^{\rm P}\left(a\right)=\frac{\pi}{2}\frac{{\rm sgn}\left(T_{\rm P}\cos{\theta}-a\right)}{\sqrt{a^{2}-2T_{\rm P}\cos{\theta}a+T_{\rm P}^{2}-\sin^{2}{\theta}}},
&& U_{2}^{\rm P}\left(a\right)=\frac{\pi}{2}\frac{\cos{\frac{\theta_{1}+\theta_{2}}{2}}}{a\sqrt{A_{1}A_{2}}},\\
&U_{3}^{\rm P}\left(a\right)=\frac{\pi}{2}\frac{\sin{\frac{\theta_{1}+\theta_{2}}{2}}}{\sqrt{A_{1}A_{2}}},
&& U_{4}^{\rm P}=\frac{\pi}{2},\\
\label{Ulast}&U_{5}^{\rm P}=\frac{\pi}{2}T_{\rm P}\cos{\theta},
&& U_{6}^{\rm P}
=\frac{\pi}{2}T_{\rm P}^{2}\cos^{2}{\theta}-\frac{\pi}{4}(T_{\rm P}^{2}-1)\sin^{2}{\theta},
\end{align}
where
${\rm sgn}(\cdot)$ represent the sign function,
$A_{1}A_{2}e^{{\rm i}\left(\theta_{1}+\theta_{2}\right)}=a^{2}+\sin^{2}{\theta}-T_{\rm P}^{2}+2{\rm i}T_{\rm P}\cos{\theta}a$.
\subsection{Calculation of $V^{\rm P}$}
In order to employ standard elliptic integrals, we proceed to introduce two parameters $\xi_{1}$ and
$\xi_{2}$ ($<\xi_1$) that satisfy
\begin{align}
\label{eqn4}\xi^{2}-\frac{T_{\rm P}^{2}+\cos^{2}{\theta}-k^{2}}{T_{\rm P}\cos{\theta}}\xi+1-k^{2}=0.
\end{align}
It is easy to show that $\xi_{2}<0<T_{\rm P}\cos{\theta}<\xi_{1}$ when $T_{\rm P}>1$.
Letting $B=\frac{\xi_{2}C-\xi_{1}}{C-1}$ (see Armitage and Eberlein, 
{\color{blue}2006}, p.220), we have that
\begin{align*}
\frac{\mathrm{d}B}{\sqrt{Q_{2}\left(B\right)}}=\frac{1}{\sqrt{\xi_{2}(\xi_{2}-T_{\rm P}\cos{\theta})}}\frac{\mathrm{d}C}{\sqrt{\left(C^{2}+C_{1}^{2}\right)\left(C^{2}+C_{2}^{2}\right)}},
\end{align*}
where
$C^{\rm P}_{1}=\sqrt{-\xi_{1}/\xi_{2}}$ and $C^{\rm P}_{2}=\sqrt{\frac{\xi_{1}-T_{\rm P}\cos{\theta}}{T_{\rm P}\cos{\theta}-\xi_{2}}}$.
By the substitution, we transform the integral path from a straight line parallel to imaginary axis (Fig. 2 (a)) to an arc called $\Gamma_{1}$ (Fig. 2 (b)),  with starting point at $C_{1}\left(-\frac{\xi_{1}-T_{\rm P}\cos{\theta}}{T_{\rm P}\cos{\theta}-\xi_{2}},0\right)$ and ending point $C_{2}\left(0,\sqrt{\frac{\xi_{1}-T_{\rm P}\cos{\theta}}{T_{\rm P}\cos{\theta}-\xi_{2}}}{\rm i}\right)$. Based on the residue theorem, the integral path $\Gamma_1$ can be changed to a new one $\Gamma_2$, which travels first along the real axis and then along the imaginary axis. For the integral $V_{1}^{\rm P}(a)$, we must examine whether there is a pole between $C_{1}$ and zero, which can contribute a value with a half of its residue. For other integrals there is no pole located in the area encircled by $\Gamma_{1}$ and $\Gamma_{2}$. The integral along the real axis has no contribution to the result, and noticing that $C_{2}$ is a branch point, the integral along the imaginary axis will generate some complete elliptic integrals.
\begin{figure}
\centering
\subfigure[]{
\begin{minipage}{.48\textwidth}
\centering
\includegraphics[scale=0.55]{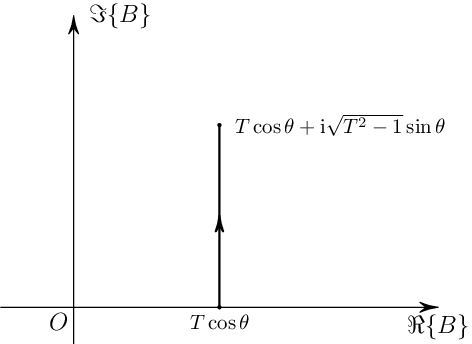}
\end{minipage}
}
\subfigure[]{
\begin{minipage}{.48\textwidth}
\centering
\includegraphics[scale=0.55]{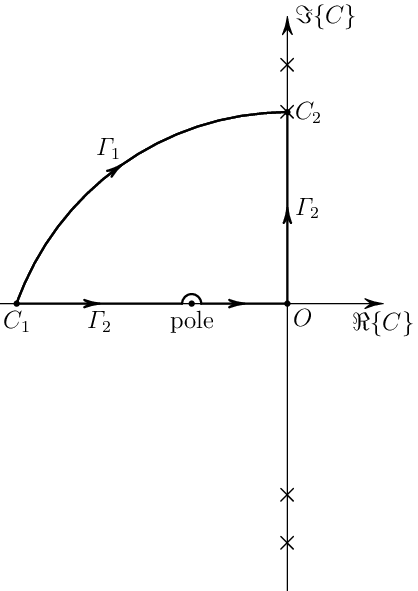}
\end{minipage}
}
\caption{(a) Integral path by variable $B$. The real axis represents the real part of $B$ and the imaginary axis the imaginary part of $B$. (b) Integral path by variable $C$, we change the path from $\Gamma_{1}$ to $\Gamma_{2}$, $\times$ means the branch point. The point on the real axis means a pole.}
\end{figure}

Three basic types of standard elliptic integrals are defined as
\begin{align*}
K\left(\tau\right)&=\int_{0}^{1}\frac{\mathrm{d}x}{\sqrt{1-x^{2}}\sqrt{1-\tau^{2}x^{2}}},\\
E\left(\tau\right)&=\int_{0}^{1}\frac{\sqrt{1-\tau^{2}x^{2}}}{\sqrt{1-x^{2}}}\mathrm{d}x,\\
\varPi\left(\tau,c\right)&=\int_{0}^{1}\frac{1}{1-cx^{2}}\frac{\mathrm{d}x}{\sqrt{1-x^{2}}\sqrt{1-\tau^{2}x^{2}}},
\end{align*}
where $0<\tau<1$ and $c<1$.
All the types can be easily obtained by Matlab. $K\left(\tau\right)$ and $E\left(\tau\right)$ can be evaluated with the command ``ellipke" and $\varPi\left(\tau,c\right)$ with ``ellipticPi".
After discarding all the real parts of these integrals, the final results can be simplified as a series of elementary algebraic expressions and three kinds of standard elliptic integrals.
As for $V_{6}^{\rm P}$ and $V_{7}^{\rm P}$, integrals with the form $\int\left(1- C^{2}\right)^{-m}\left[\left(C^{2}+C_{1}^{2}\right)\left(C^{2}+C_{2}^{2}\right)\right]^{-1/2}\mathrm{d}t$ should be derived, in which $m$ is a positive integral. By differentiating the expressions of the form $C(1-C^{2})^{1-m}\big[(C^{2}+C_{1}^{2})(C^{2}+C_{2}^{2})\big]^{1/2}$, we obtain the reduction formulae for the above integrals, and finally the integrals can be expressed in terms of three standerd elliptic integrals. More details can be found in Armitage and Eberlein ({\color{blue}2006}).
\subsection{Results of $V^{\rm P}$}
Through tedious derivations, all the integrals $V_{i}^{\rm P} $ $(i=1,2,\cdots,7)$ can be solved exactly as
\begin{align}
V_{1}^{\rm P}(a)
=&-\left[H\left(a-T_{\rm P}\cos{\theta}\right)-H\left(a-\xi_{1}\right)\right]\frac{1}{\sqrt{a^{2}+k^{2}-1}}\frac{\pi}{\sqrt{a^{2}-2T_{\rm P}\cos{\theta}a+T_{\rm P}^{2}-\sin^{2}{\theta}}} \nonumber\\
\label{Vfirst}&+\frac{1}{\xi_{2}-a}M_{\rm P}K_{\rm P}+\frac{\xi_{2}-\xi_{1}}{\left(\xi_{1}-a\right)\left(\xi_{2}-a\right)}M_{\rm P}\Pi_{\rm P}\left(-\left[\frac{\left(\xi_{2}-a\right)C^{\rm P}_{2}}{\xi_{1}-a}\right]^{2}\right),\\
V^{\rm P}_{2}(a)
=&\frac{M_{\rm P}}{a^{2}+\xi_{2}^{2}}\left\{2\left(\xi_{1}-\xi_{2}\right)\Re\left[\frac{m_{1}}{n_{1}}\Pi_{\rm P}\left(-\left(C^{\rm P}_{2}\right)^{2}\left(a^{2}+\xi_{2}^{2}\right)^{2}/n_{1}\right)\right]+K_{\rm P}\right\},\\
V^{\rm P}_{3}(a)
=&\frac{M_{\rm P}}{a^{2}+\xi_{2}^{2}}\left\{-2\left(\xi_{1}-\xi_{2}\right)a\Im\left[\frac{m_{1}}{n_{1}}\Pi_{\rm P}\left(-\left(C^{\rm P}_{2}\right)^{2}\left(a^{2}+\xi_{2}^{2}\right)^{2}/n_{1}\right)\right]+\xi_{2}K_{\rm P}\right\},\\
V_{4}^{\rm P}=&M_{\rm P}K_{\rm P},\\
V_{5}^{\rm P}=&M_{\rm P}\left[\xi_{2}K_{\rm P}+\left(\xi_{1}-\xi_{2}\right)\Pi_{\rm P}\left(-\left(C_{2}^{\rm P}\right)^{2}\right)\right],\\
V_{6}^{\rm P}
=&M_{\rm P}\left[T_{\rm P}\cos{\theta}\xi_{2}K_{\rm P}+\xi_{1}\left(T_{\rm P}\cos{\theta}-\xi_{2}\right)\left(E_{\rm P}-K_{\rm P}\right)+T_{\rm P}\cos{\theta}\left(\xi_{1}-\xi_{2}\right)\Pi_{\rm P}\left(-\left(C^{\rm P}_{2}\right)^{2}\right)\right],\\
V_{7}^{\rm P}
=&\frac{M_{\rm P}}{2}\Big[\xi_{2}\left(3T_{\rm P}^{2}\cos^{2}{\theta}-T_{\rm P}^{2}+2-k^{2}-\cos^{2}{\theta}-T_{\rm P}\cos{\theta}\xi_{1}\right)K_{\rm P} \nonumber\\
&+\left(\xi_{1}-\xi_{2}\right)\left(3T_{\rm P}^{2}\cos^{2}{\theta}-T_{\rm P}^{2}+2-k^{2}-\cos^{2}{\theta}\right)\Pi_{\rm P}\left(-\left(C^{\rm P}_{2}\right)^{2}\right)\nonumber\\
\label{Vlast}&+3T_{\rm P}\cos{\theta}\xi_{1}\left(T_{\rm P}\cos{\theta}-\xi_{2}\right)\left(E_{\rm P}-K_{\rm P}\right)\Big],
\end{align}
where
\begin{align*}
&M_{\rm P}=\frac{1}{\sqrt{\xi_{1}(T_{\rm P}\cos{\theta}-\xi_{2})}}, && \tau_{\rm P}=\sqrt{\frac{\xi_{2}(\xi_{1}-T_{\rm P}\cos{\theta})}{\xi_{1}(\xi_{2}-T_{\rm P}\cos{\theta})}}, && C^{\rm P}_{2}=\sqrt{\frac{\xi_{1}-T_{\rm P}\cos{\theta}}{T_{\rm P}\cos{\theta}-\xi_{2}}},\\
&K(\tau_{\rm P})=K_{\rm P}, && E(\tau_{\rm P})=E_{\rm P}, && \Pi(\tau_{\rm P}, \cdot)=\Pi_{\rm P}(\cdot),
\end{align*}
\begin{align*}
&m_{1}=\frac{1}{2}(\xi_{1}+\xi_{2})\left(a^{2}+\xi_{2}^{2}\right)-2\xi_{2}\left(a^{2}+\xi_{1}\xi_{2}\right)-{\rm i}\left[\frac{a^{2}+\xi_{1}\xi_{2}}{2a}\left(a^{2}-\xi_{2}^{2}\right)+\xi_{2}\left(\xi_{1}-\xi_{2}\right)a\right],\\
&n_{1}=2\left(a^{2}+\xi_{1}\xi_{2}\right)^{2}-\left(a^{2}+\xi_{1}^{2}\right)\left(a^{2}+\xi_{2}^{2}\right)+2{\rm i}a\left(a^{2}+\xi_{1}\xi_{2}\right)\left(\xi_{1}-\xi_{2}\right).
\end{align*}
It is essential to point out that the expressions of $V_{i}^{\rm P}\ (i=1,2,\cdots,7)$ only contain the parameters defined in the section  ``Notation and definitions'' and $\xi_{1}$ and $\xi_{2}$ defined in the equation ({\ref{eqn4}}).

\subsection{Summary}
So far, we have succeeded in obtaining the closed-form of $\ts{G}^{\rm P}$.
The procedures are so elaborate that an explicit guide to our formulae should be listed below to allow readers to implement our results easily using Matlab:
\begin{enumerate}
\item Use the command ``residue'' to obtain the coefficient $u_{ij,k}^{\rm P}\ (k=1,2,\cdots,9)$ with the individual components of the matrix $\ts{M}^{\rm P}$(equations ({\ref{Mfirst})--(\ref{Mlast}})) and  $R^{\rm P}(B)$ (equation ({\ref{RP}})). The inputs to ``residue'' are vectors of coefficients of the polynomials in the numerator and the denominator, i.e. $M_{ij}^{\rm P}(B)$ and $R^{\rm P}(B)$.  The outputs are the residues ($u_{ij,k}^{\rm P}\ (k=1,2,\cdots,6)$), the poles ($a_{1}^{\rm P}$, $a_{2}^{\rm P}$ and $a_{3}^{\rm P}$) and the quotient ($u_{ij,k}^{\rm P}\ (k=7,8,9)$).
\item Use the command ``residue'' to obtain the coefficient $v_{ij,k}^{\rm P}\ (k=1,2,\cdots,10)$ with every components of the matrix $\ts{N}^{\rm P}$(equations ({\ref{Nfirst})--(\ref{Nlast}})) and  $R^{\rm P}(B)$. The inputs to ``residue'' are vectors of coefficients of $N_{ij}^{\rm P}(B)$ and $R^{\rm P}(B)$.  The outputs are the residues ($v_{ij,k}^{\rm P}\ (k=1,2,\cdots,6)$), the poles ($a_{1}^{\rm P}$, $a_{2}^{\rm P}$ and $a_{3}^{\rm P}$) and the quotient ($v_{ij,k}^{\rm P}\ (k=7,\cdots,10)$).
\item Express the results of $U_{i}^{\rm P}\ (i=1,2,\cdots,6)$(equations ({\ref{Ufirst})--(\ref{Ulast}})). All the parameters have been defined in the Section 2.
\item Express the results of $V_{i}^{\rm P}\ (i=1,2,\cdots,7)$(equations ({\ref{Vfirst})--(\ref{Vlast}})). $\xi_{1}$ and $\xi_{2}$ could be obtained by the equation (32).
\item Combine these results by the equation ({\ref{GP}}).
\end{enumerate}
All the code has been provided as additional material on the journal site. We should pay particular attention to $U^{\rm P}_{2}(a)$, $U^{\rm P}_{3}(a)$, $V^{\rm P}_{2}(a)$ and $V^{\rm P}_{3}(a)$, which contain the actual root of the Rayleigh function. These expressions contain all the properties of the Rayleigh wave. The numerical examples corroborate the argument that these expressions generate the Rayleigh wave significantly when the depth of source is small compared with the epicentral distance, while other terms have nothing to do with the Rayleigh wave. We separate the expressions of the Rayleigh wave from the whole Green's function, which sheds light on the contribution of the Rayleigh waves to the motion observed in a homogeneous half-space medium.
\section{Calculation of $\ts{G}^{\rm S}$}
Equation ({\ref{eqn2}}) can be handled with a similar manner as for $\ts{G}^{\rm P}$. With substitution $B=\beta\eta_{_\beta}$, we obtain that
\begin{align}
\ts{G}^{\rm S}=\frac{1}{\pi^{2}\mu r}\frac{\partial}{\partial t}\left(\Im\int_{B_{2}}^{T^{\rm S}_{\rm upper}}H\left(t-t_{_{\rm S}}\right)\frac{\ts{M}^{\rm S}(B)}{R^{\rm S}(B)}\frac{\mathrm{d}B}{\sqrt{Q^{\rm S}_{1}\left(B\right)}}+\Im\int_{B_{2}}^{T^{\rm S}_{\rm upper}}H(t-t_{_{\rm S}})\frac{\ts{N}^{\rm S}(B)}{R^{\rm S}(B)}\frac{\mathrm{d}B}{\sqrt{Q^{\rm S}_{2}\left(B\right)}}\right).
\end{align}
where 
\begin{align*}
&T^{\rm S}_{\rm upper}=T_{\rm S}\cos{\theta}+{\rm i}\sqrt{T_{\rm S}^{2}-1}\sin{\theta}, &&
R^{\rm S}(B)=(2B^{2}-1)^{4}-16B^{2}(B^{2}+k^{-2}-1)(B^{2}-1)^{2},\\
&Q^{\rm S}_{1}(B)=B^{2}-2T_{\rm S}\cos{\theta}B+T_{\rm S}^{2}-\sin^{2}{\theta}, && 
Q^{\rm S}_{2}(B)=(B^{2}+k^{-2}-1)(B^{2}-2T_{\rm S}\cos{\theta}B+T_{\rm S}^{2}-\sin^{2}{\theta}),\\
&B_{2}=\max\left(T_{\rm S}\cos{\theta},\sqrt{1-k^{-2}}\right).
\end{align*}
$R^{\rm S}(B^2)$ is equivalent to the Rayleigh function $\sigma(y)$ noticing that $y=1-B^2$. So $R^{\rm S}(B^2)$ has a negative root $-(a^{\rm S}_{3})^2=1-y_{3}$ (actual zero) and two positive roots $(a^{\rm S}_{1})^2=1-y_{1}$ and $(a^{\rm S}_{2})^2=1-y_{2}$. $a^{\rm S}_{1}$, $a^{\rm S}_{2}$ and $a^{\rm S}_{3}$ are all positive numbers.
The expressions for the individual elements of the three-by-three matrics $\ts{M}^{\rm S}$ and $\ts{N}^{\rm S}$ are as follows:
\begin{align}
M^{\rm S}_{11}&=I^{\rm S}_{1}\cos^{2}{\phi}+I^{\rm S}_{2}\sin^{2}{\phi},\\
M^{\rm S}_{22}&=I^{\rm S}_{1}\sin^{2}{\phi}+I^{\rm S}_{2}\cos^{2}{\phi},\\
M^{\rm S}_{12}&=M^{\rm S}_{21}=(I^{\rm S}_{1}-I^{\rm S}_{2})\sin{\phi}\cos{\phi},\\
M^{\rm S}_{13}&=-\frac{1}{\sin{\theta}}B(2B^{2}-1)^{3}(\cos{\theta}B-T_{\rm S})\cos{\phi},\\
M^{\rm S}_{23}&=-\frac{1}{\sin{\theta}}B(2B^{2}-1)^{3}(\cos{\theta}B-T_{\rm S})\sin{\phi},\\
M^{\rm S}_{31}&=-\frac{8}{\sin{\theta}}B^{3}(B^{2}-1)(B^{2}+k^{-2}-1)(\cos{\theta}B-T_{\rm S})\cos{\phi},\\
M^{\rm S}_{32}&=-\frac{8}{\sin{\theta}}B^{3}(B^{2}-1)(B^{2}+k^{-2}-1)(\cos{\theta}B-T_{\rm S})\sin{\phi},\\
M^{\rm S}_{33}&=-8B^{2}(B^{2}-1)^{2}(B^{2}+k^{-2}-1),
\end{align}
in which
\begin{align*}
I^{\rm S}_{1}&=R^{\rm S}\left(B\right)+\frac{1}{\sin^{2}{\theta}}\left(\left(2B^{2}-1\right)^{3}-16B^{2}\left(B^{2}+k^{-2}-1\right)\left(B^{2}-1\right)\right)\left(\cos{\theta}B-T_{\rm S}\right)^{2},\\
I_{2}&=R^{\rm S}\left(B\right)+\frac{8}{\sin^{2}{\theta}}\left(\left(2B^{2}-1\right)^{3}-16B^{2}\left(B^{2}+k^{-2}-1\right)\left(B^{2}-1\right)\right)\left(B^{2}-2T_{\rm S}\cos{\theta}B+T_{\rm S}^{2}-\sin^{2}{\theta}\right),
\end{align*}
and
\begin{align}
N^{\rm S}_{11}&=I^{\rm S}_{3}\cos^{2}{\phi}+I^{\rm S}_{4}\sin^{2}{\phi},\\
N^{\rm S}_{22}&=I^{\rm S}_{3}\sin^{2}{\phi}+I^{\rm S}_{4}\cos^{2}{\phi},\\
N^{\rm S}_{12}&=N^{\rm S}_{21}=(I^{\rm S}_{3}-I^{\rm S}_{4})\sin{\phi}\cos{\phi},\\
N^{\rm S}_{13}&=2(B^{2}-1)I^{\rm S}_{5}\cos{\phi},\\
N^{\rm S}_{23}&=2(B^{2}-1)I^{\rm S}_{5}\sin{\phi},\\
N^{\rm S}_{31}&=(2B^{2}-1)I^{\rm S}_{5}\cos{\phi},\\
N^{\rm S}_{32}&=(2B^{2}-1)I^{\rm S}_{5}\sin{\phi},\\
N^{\rm S}_{33}&=-2B(B^{2}+k^{-2}-1)(2B^{2}-1)^{2}(B^{2}-1),
\end{align}
in which
\begin{align*}
I^{\rm S}_{3}&=-\frac{4}{\sin^{2}{\theta}}B^{3}(2B^{2}-1)(B^{2}+k^{-2}-1)(\cos{\theta}B-T_{\rm S})^{2},\\
I^{\rm S}_{4}&=\frac{4}{\sin^{2}{\theta}}B^{3}(2B^{2}-1)(B^{2}+k^{-2}-1)(B^{2}-2T_{\rm S}\cos{\theta}B+T_{\rm S}^{2}-\sin^{2}{\theta}),\\
I^{\rm S}_{5}&=-\frac{2}{\sin{\theta}}B^{2}(B^{2}+k^{-2}-1)(2B^{2}-1)(\cos{\theta}B-T_{\rm S}).
\end{align*}
The final results for $\ts{G}^{\rm S}$ are obtained in a way similar as those for $\ts{G}^{\rm P}$, so we just state the key points of the procedures. 
Introducing $\xi_{1}$ and $\xi_{2}$  ($<\xi_1$) that satisfy the quadratic equation
\begin{align}
\label{eqn5}\xi^{2}-\frac{T_{\rm S}^{2}+\cos^{2}{\theta}-k^{-2}}{T_{\rm S}\cos{\theta}}\xi+1-k^{-2}=0.
\end{align}
When $T_{\rm S}>1$, the inequality below is valid,
\begin{align*}
0<\xi_{2}<T_{\rm S}\cos{\theta}<\xi_{1}.
\end{align*}
And we expand $\frac{M^{\rm S}_{ij}(B)}{R^{\rm S}(B)}$ and $\frac{N^{\rm S}_{ij}(B)}{R^{\rm S}(B)}$ into the monomials and the partial fractions as
\begin{align}
\frac{M^{\rm S}_{ij}(B)}{R^{\rm S}(B)}=&\frac{u^{\rm S}_{ij,1}}{B-a^{\rm S}_{1}}+
\frac{u^{\rm S}_{ij,2}}{B+a^{\rm S}_{1}}+\frac{u^{\rm S}_{ij,3}}{B-a^{\rm S}_{2}}+\frac{u^{\rm S}_{ij,4}}{B+a^{\rm S}_{2}}+\frac{u^{\rm S}_{ij,5}}{B^{2}+\left(a^{\rm S}_{3}\right)^{2}}+\frac{u^{\rm S}_{ij,6}B}{B^{2}+\left(a^{\rm S}_{3}\right)^{2}}\nonumber \\
&+u^{\rm S}_{ij,7}+u^{\rm S}_{ij,8}B+u^{\rm S}_{ij,9}B^{2},\\
\frac{N^{\rm S}_{ij}(B)}{R^{\rm S}(B)}=&\frac{v^{\rm S}_{ij,1}}{B-a^{\rm S}_{1}}+
\frac{v^{\rm S}_{ij,2}}{B+a^{\rm S}_{1}}+\frac{v^{\rm S}_{ij,3}}{B-a^{\rm S}_{2}}+\frac{v^{\rm S}_{ij,4}}{B+a^{\rm S}_{2}}+\frac{v^{\rm S}_{ij,5}}{B^{2}+\left(a^{\rm S}_{3}\right)^{2}}+\frac{v^{\rm S}_{ij,6}B}{B^{2}+\left(a^{\rm S}_{3}\right)^{2}}\nonumber \\
\label{gssp}&+v^{\rm S}_{ij,7}+v^{\rm S}_{ij,8}B+v^{\rm S}_{ij,9}B^{2}+v^{\rm S}_{ij,10}B^{3},
\end{align}
The final result of $G^{\rm S}_{ij}$ can be obtained as
\begin{align}
G^{\rm S}_{ij}=&\frac{1}{\pi^{2}\mu r}\frac{\partial}{\partial t}\Big\{H\left(t-t_{_{\rm S}}\right)\Big[u^{\rm S}_{ij,1}U^{\rm S}_{1}\left(a^{\rm S}_{1}\right)+u^{\rm S}_{ij,2}U^{\rm S}_{1}\left(-a^{\rm S}_{1}\right)+u^{\rm S}_{ij,3}U^{\rm S}_{1}\left(a^{\rm S}_{2}\right)+u^{\rm S}_{ij,4}U^{\rm S}_{1}\left(-a^{\rm S}_{2}\right)\nonumber
\\&+u^{\rm S}_{ij,5}U^{\rm S}_{2}\left(a^{\rm S}_{3}\right)+u^{\rm S}_{ij,6}U^{\rm S}_{3}\left(a^{\rm S}_{3}\right)+u^{\rm S}_{ij,7}U^{\rm S}_{4}+u^{\rm S}_{ij,8}U^{\rm S}_{5}+u^{\rm S}_{ij,9}U^{\rm S}_{6}\Big]\Big\}\nonumber
\\&+\frac{1}{\pi^{2}\mu r}\frac{\partial}{\partial t}\Big\{H\left(t-t_{_{\rm S}}\right)\Big[v^{\rm S}_{ij,1}V^{\rm S}_{1}\left(a^{\rm S}_{1}\right)+v^{\rm S}_{ij,2}V^{\rm S}_{1}\left(-a^{\rm S}_{1}\right)+v^{\rm S}_{ij,3}V^{\rm S}_{1}\left(a^{\rm S}_{2}\right)+v^{\rm S}_{ij,4}V^{\rm S}_{1}\left(-a^{\rm S}_{2}\right)\nonumber
\\&+v^{\rm S}_{ij,5}V^{\rm S}_{2}\left(a^{\rm S}_{3}\right)+v^{\rm S}_{ij,6}V^{\rm S}_{3}\left(a^{\rm S}_{3}\right)+v^{\rm S}_{ij,7}V^{\rm S}_{4}+v^{\rm S}_{ij,8}V^{\rm S}_{5}+v^{\rm S}_{ij,9}V^{\rm S}_{6}+v^{\rm S}_{ij,10}V^{\rm S}_{7}\Big]\Big\},
\end{align}
where
\begin{align}
&V^{\rm S}_{1}(a)
=\frac{M_{\rm S}}{\xi_{2}-a}\left[\frac{(\xi_{1}-a)(\xi_{2}-T_{\rm S}\cos{\theta})}{a^{2}+T_{\rm S}^{2}-2aT_{\rm S}\cos{\theta}-\sin^{2}{\theta}}\Pi_{\rm S}\left(\frac{\left(C^{\rm S}_{2}\right)^{2}\left(\xi_{2}-a\right)^{2}}{\left(C^{\rm S}_{2}\right)^{2}\left(\xi_{2}-a\right)^{2}+\left(\xi_{1}-a\right)^{2}}\right)+K_{\rm S}\right]\nonumber\\
&-\left(H\left(a-T_{\rm S}\cos{\theta}\right)-H\left(a-\xi_{1}\right)\right)\frac{\pi}{\sqrt{a^{2}+k^{-2}-1}\sqrt{a^{2}-2aT_{\rm S}\cos{\theta}+T_{\rm S}^{2}-\sin^{2}{\theta}}},
\end{align}
\begin{align}
V^{\rm S}_{2}(a)
=&\frac{M_{\rm S}}{a^{2}+\xi_{2}^{2}}\left\{2(\xi_{1}-\xi_{2})\Re\big[\frac{m_{1}}{n_{1}}\Pi_{\rm S}\left(\left(C^{\rm S}_{2}\right)^{2}\left(a^{2}+\xi_{2}^{2}\right)^{2}/n_{1}\right)\big]+K_{\rm S}\right\},\\
V^{\rm S}_{3}(a)
=&\frac{M_{\rm S}}{a^{2}+\xi_{2}^{2}}\left\{-2(\xi_{1}-\xi_{2})a\Im\left[\frac{m_{1}}{n_{1}}\Pi_{\rm S}\left(\left(C^{\rm S}_{2}\right)^{2}\left(a^{2}+\xi_{2}^{2}\right)^{2}/n_{1}\right)\right]+\xi_{2}K_{\rm S}\right\},\\
V^{\rm S}_{4}=&M_{\rm S}K_{\rm S},\\
V^{\rm S}_{5}
=&M_{\rm S}\left[\xi_{2}K_{\rm S}+\left(T_{\rm S}\cos{\theta}-\xi_{2}\right)\Pi_{\rm S}\left(\frac{\xi_{1}-T_{\rm S}\cos{\theta}}{\xi_{1}-\xi_{2}}\right)\right],\\
V^{\rm S}_{6}
=&M_{\rm S}\Big[T_{\rm S}\cos{\theta}\left(\xi_{1}-\xi_{2}\right)E_{\rm S}+\left(\xi_{1}\xi_{2}-\left(\xi_{1}-\xi_{2}\right)T_{\rm S}\cos{\theta}\right)K_{\rm S}\nonumber\\
&+\left(T_{\rm S}\cos{\theta}-\xi_{2}\right)T_{\rm S}\cos{\theta}\Pi_{\rm S}\left(\frac{\xi_{1}-T_{\rm S}\cos{\theta}}{\xi_{1}-\xi_{2}}\right)\Big],\\
V^{\rm S}_{7}
=&\frac{M_{\rm S}}{2}\Big[\xi_{2}\left(3T_{\rm S}^{2}\cos^{2}{\theta}-T_{\rm S}^{2}+2-k^{-2}-\cos^{2}{\theta}-T_{\rm S}\cos{\theta}\xi_{1}\right)K_{\rm S}\nonumber\\&-3T_{\rm S}\cos{\theta}\xi_{1}\left(T_{\rm S}\cos{\theta}-\xi_{2}\right)K_{\rm S}+3T_{\rm S}^{2}\cos^{2}{\theta}\left(\xi_{1}-\xi_{2}\right)E_{\rm S}\nonumber\\
&+\left(T_{\rm S}\cos{\theta}-\xi_{2}\right)\left(3T_{\rm S}^{2}\cos^{2}{\theta}-T_{\rm S}^{2}+2-k^{-2}-\cos^{2}{\theta}\right)\Pi_{\rm S}\left(\frac{\xi_{1}-T_{\rm S}\cos{\theta}}{\xi_{1}-\xi_{2}}\right)\Big],
\end{align}
where
\begin{align*}
&C^{\rm S}_{2}=\sqrt{\frac{\xi_{1}-T_{\rm S}\cos{\theta}}{T_{\rm S}\cos{\theta}-\xi_{2}}}, && M_{\rm S}=\frac{1}{\sqrt{T_{\rm S}\cos{\theta}(\xi_{1}-\xi_{2})}}, && \tau_{\rm S}=\sqrt{\frac{\xi_{2}(\xi_{1}-T\cos{\theta})}{T_{\rm S}\cos{\theta}(\xi_{1}-\xi_{2})}}, \\
&K(\tau_{\rm S})=K_{\rm S}, && E(\tau_{\rm S})=E_{\rm S}, && \Pi(\tau_{\rm S},\cdot)=\Pi_{\rm S}(\cdot),\\
\end{align*}
and
\begin{align*}
&m_{1}=\frac{1}{2}\left(\xi_{1}+\xi_{2}\right)\left(a^{2}+\xi_{2}^{2}\right)-2\xi_{2}\left(a^{2}+\xi_{1}\xi_{2}\right)-{\rm i}\left[\frac{a^{2}+\xi_{1}\xi_{2}}{2a}\left(a^{2}-\xi_{2}^{2}\right)+\xi_{2}\left(\xi_{1}-\xi_{2}\right)a\right],\\
&n_{1}=\left(C^{\rm S}_{2}\right)^{2}\left(a^{2}+\xi_{2}^{2}\right)^{2}+2\left(a^{2}+\xi_{1}\xi_{2}\right)^{2}-\left(a^{2}+\xi_{1}^{2}\right)\left(a^{2}+\xi_{2}^{2}\right)+2{\rm i}a\left(a^{2}+\xi_{1}\xi_{2}\right)\left(\xi_{1}-\xi_{2}\right).
\end{align*} 
For every $i$ $(i=1,2,\cdots,6)$, the result of $U^{\rm S}_{i}$ is identical to $U^{\rm P}_{i}$, except changing $T_{\rm P}$ to $T_{\rm S}$.
\section{Calculation of $\ts{G}^{\rm S\text{-}P}$}
With the same substitution $B=\beta\eta_{_\beta}$, equation ({\ref{eqn3}}) can be transformed as
\begin{align}
\ts{G}^{\rm S\text{-}P}=\frac{H\left(\sin{\theta}-k^{-1}\right)}{\pi^{2}\mu r}\frac{\partial}{\partial t}\Im\int_{T^{\rm S}_{\rm lower}}^{\sqrt{1-k^{-2}}}\left(H\left(t-t_{_{\rm S\text{-}P}}\right)-H\left(t-t_{\rm S}\right)\right)\frac{\ts{N}^{\rm S}\left(B\right)}{R^{\rm S}\left(B\right)}\frac{\mathrm{d}B}{\sqrt{Q^{\rm S}_{2}\left(B\right)}},
\end{align}
where 
$T^{\rm S}_{\rm lower}=T_{\rm S}\cos{\theta}+\sqrt{1-T_{\rm S}^{2}}\sin{\theta}$.
 Other signals have the same definitions as $\ts{G}^{\rm S}$. 
$\xi_{1}$ and $\xi_{2}$ are two roots of the quadratic equation,
\begin{align}
\label{eqn6}\xi^{2}-\frac{T_{\rm S}^{2}+\cos^{2}{\theta}-k^{-2}}{T_{\rm S}\cos{\theta}}\xi+1-k^{-2}=0.
\end{align}
It is noted that $0<T_{\rm S}\cos{\theta}<\xi_{2}<\xi_{1}$, when $T_{\rm S}<1$.
With equation (\ref{gssp}), the final result of $G^{\rm S\text{-}P}_{ij}$ is obtained as
\begin{align}
G^{\rm S\text{-}P}_{ij}=&\frac{H\left(\sin{\theta}-k^{-1}\right)}{\pi^{2}\mu r}\frac{\partial}{\partial t}\Big\{\left(H\left(t-t_{\rm S\text{-}P}\right)-H\left(t-t_{_{\rm S}}\right)\right)
\Big(v^{\rm S}_{ij,1}V^{\rm S\text{-}P}_{1}(a^{\rm S}_{1})\nonumber\\&+v^{\rm S}_{ij,2}V^{\rm S\text{-}P}_{1}(-a^{\rm S}_{1})+v^{\rm S}_{ij,3}V^{\rm S\text{-}P}_{1}(a^{\rm S}_{2})+v^{\rm S}_{ij,4}V^{\rm S\text{-}P}_{1}(-a^{\rm S}_{2})
+v^{\rm S}_{ij,5}V^{\rm S\text{-}P}_{2}(a^{\rm S}_{3})\nonumber\\&+v^{\rm S}_{ij,6}V^{\rm S\text{-}P}_{3}(a^{\rm S}_{3})+v^{\rm S}_{ij,7}V^{\rm S\text{-}P}_{4}+v^{\rm S}_{ij,8}V^{\rm S\text{-}P}_{5}+v^{\rm S}_{ij,9}V^{\rm S\text{-}P}_{6}+v^{\rm S}_{ij,10}V^{\rm S\text{-}P}_{7}\Big)\Big\},
\end{align}
where
\begin{small}
\begin{align}
V^{\rm S\text{-}P}_{1}\left(a\right)=&\frac{M_{\rm S\text{-}P}}{\xi_{2}-a}\left[\frac{\left(\xi_{2}-T_{\rm S}\cos{\theta}\right)\left(\xi_{1}-a\right)}{a^{2}-2T_{\rm S}\cos{\theta}a+T_{\rm S}^{2}-\sin^{2}{\theta}}\Pi_{\rm S\text{-}P}\left(c\right)+K_{\rm S\text{-}P}\right]\nonumber\\
&-\frac{\pi}{2\sqrt{\left(a^{2}-1+k^{-2}\right)\left(a^{2}-2T_{\rm S}\cos{\theta}a+T_{\rm S}^{2}-\sin^{2}{\theta}\right)}},\\
V^{\rm S\text{-}P}_{2}\left(a\right)=&M_{\rm S\text{-}P}\left(\frac{T_{\rm S}\cos{\theta}-\xi_{2}}{a}\Im\{y_{1}\}+\frac{K_{\rm S\text{-}P}}{a^{2}+\xi_{2}^{2}}\right)-\frac{\pi}{2a\sqrt{a^{2}+1-k^{-2}}}\frac{\sin{\frac{\theta_{1}}{2}}}{\sqrt{A_{1}}},\\
V^{\rm S\text{-}P}_{3}\left(a\right)=&M_{\rm S\text{-}P}\left(\left(T_{\rm S}\cos{\theta}-\xi_{2}\right)\Re\{y_{1}\}+\frac{\xi_{2}}{a^{2}+\xi_{2}^{2}}K_{\rm S\text{-}P}\right)+\frac{\pi}{2\sqrt{a^{2}+1-k^{-2}}}\frac{\cos{\frac{\theta_{1}}{2}}}{\sqrt{A_{1}}},\\
V^{\rm S\text{-}P}_{4}=&M_{\rm S\text{-}P}K_{\rm S\text{-}P},\\
V^{\rm S\text{-}P}_{5}=&M_{\rm S\text{-}P}\left[\xi_{2}K_{\rm S\text{-}P}+\left(T_{\rm S}\cos{\theta}-\xi_{2}\right)\Pi_{\rm S\text{-}P}\left(T_{\rm S}\cos{\theta}/\xi_{2}\right)\right]+\frac{\pi}{2},\\
V^{\rm S\text{-}P}_{6}=&M_{\rm S\text{-}P}\left[\xi_{2}T_{\rm S}\cos{\theta}K_{\rm S\text{-}P}+\xi_{2}\left(\xi_{1}-T_{\rm S}\cos{\theta}\right)E_{\rm S\text{-}P}+\left(T_{\rm S}\cos{\theta}-\xi_{2}\right)T_{\rm S}\cos{\theta}\Pi_{\rm S\text{-}P}\left(T_{\rm S}\cos{\theta}/\xi_{2}\right)\right]\nonumber \\&+\frac{\pi}{2}T_{\rm S}\cos{\theta},\\
V^{\rm S\text{-}P}_{7}=&\frac{M_{\rm S\text{-}P}}{2}\Big[\xi_{2}\left(3T_{\rm S}^{2}\cos^{2}{\theta}+2\xi_{1}\xi_{2}-\left(\xi_{1}+\xi_{2}\right)T_{\rm S}\cos{\theta}-\xi_{1}T_{\rm S}\cos{\theta}\right)K_{\rm S\text{-}P} \nonumber\\
&+\left(T_{\rm S}\cos{\theta}-\xi_{2}\right)\left(3T_{\rm S}^{2}\cos^{2}{\theta}+2\xi_{1}\xi_{2}-T_{\rm S}\cos{\theta}\left(\xi_{1}+\xi_{2}\right)\right)\Pi_{\rm S\text{-}P}\left(T_{\rm S}\cos{\theta}/\xi_{2}\right)\Big] \nonumber\\
&+
3T_{\rm S}\cos{\theta}\xi_{2}\left(\xi_{1}-T_{\rm S}\cos{\theta}\right)E_{\rm S\text{-}P}+\frac{\pi}{4}\left(3T_{\rm S}^{2}\cos^{2}{\theta}-T_{\rm S}^{2}-\cos^{2}{\theta}+2-k^{-2}\right),
\end{align}
\end{small}
where
\begin{align*}
&M_{\rm S\text{-}P}=\frac{1}{\sqrt{\xi_{2}\left(\xi_{1}-T_{\rm S}\cos{\theta}\right)}}, &&
\tau_{\rm S\text{-}P}=\sqrt{\frac{T_{\rm S}\cos{\theta}\left(\xi_{1}-\xi_{2}\right)}{\xi_{2}\left(\xi_{1}-T_{\rm S}\cos{\theta}\right)}}, &&
c=\frac{T_{\rm S}\cos{\theta}\left(\xi_{2}-a\right)^{2}}{\left(a^{2}-2T_{\rm S}\cos{\theta}a+T_{\rm S}^{2}-\sin^{2}{\theta}\right)\xi_{2}},\\
&K\left(\tau_{\rm S\text{-}P}\right)=K_{\rm S\text{-}P}, && E\left(\tau_{\rm S\text{-}P}\right)=E_{\rm S\text{-}P}, && \Pi\left(\tau_{\rm S\text{-}P}, \cdot\right)=\Pi_{\rm S\text{-}P}\left(\cdot\right),
\end{align*}
and
\begin{align*}
&y_{1}=\frac{\xi_{1}-{\rm i}a}{\xi_{2}-{\rm i}a}\frac{\Pi_{\rm S\text{-}P}\left(-c_{1}\right)}{\sin^{2}{\theta}+a^{2}-T_{\rm S}^2+2{\rm i}aT_{\rm S}\cos{\theta}},\\
&c_{1}=\frac{T_{\rm S}\cos{\theta}\left(\xi_{2}-{\rm i}a\right)^{2}}{\left(\sin^{2}{\theta}+a^{2}-T_{\rm S}^2+2{\rm i}aT\cos{\theta}\right)\xi_{2}},\\
&A_{1}e^{{\rm i}\theta_{1}}=\sin^{2}{\theta}+a^{2}-T_{\rm S}^2+2{\rm i}aT_{\rm S}\cos{\theta}.
\end{align*}

Up to now, we have expanded the original integration of $\ts{G}$ in terms of some elementary algebraic expressions and standard elliptic integrations. It should be pointed out that the spatial derivatives of the Green's function can be derived in the same manner, based on the spatial deviatives of the solutions by Johnson ({\color{blue}1974}).
\section{Numerical Results}
In this section, a few examples of computed Green's functions are compared with Johnson's ({\color{blue}1974}) results to verify the correctness of the formulae obtained in previous sections. According to our expressions, no obstacle will be faced to generate the code which has been provided as additional material on the journal site. Some essential considerations involved in the numerical calculations of the original integrations have been presented in Johnson ({\color{blue}1974}). 

In this section, we adopt the conventional symbol $G_{ij}(x_{1}, x_{2}, x_{3}, t; x'_{1}, x'_{2}, x'_{3}, \tau)$ to denote the $ij$ component of Green's function at $(x_{1}, x_{2}, x_{3}, t)$ when the source is applied at $(x'_{1}, x'_{2}, x'_{3}, \tau)$. For the sake of convenience, it is best to consider the solution which results from a source with a step time function, and the corresponding results is denoted as $\ts{G}^{\rm H}$, as Johnson (1974) did. 

All of the calculation illustated in Figs. 3--6 have a velocity of the P wave $\alpha$ of $8.00$ km ${\rm s}^{-1}$, a ratio of P to S wave velocity $k$ of $\sqrt{3}$ (Poisson solid), a density $\rho$ of $3.30$ g ${\rm cm}^{-3}$ and the coordinate of the receiver of ($10$, $0$, $0$). In Figs. 3 and 5, the coordinate of the source is ($0$, $0$, $2$) and in Figs. 4 and 6, the coordinate of the source is ($0$, $0$, $0.2$). At first we should compare our results with Johnson's ({\color{blue}1974}). In Figs. 3 (a), 4 (a), 5 (a) and 6 (a), we show the comparisons for two elements of the tensor $\ts{G}^{\rm H}$ and two different positions of the source. All of our results are identical to Johnson's ({\color{blue}1974}). This is sufficient to validate the formulae given in previous sections.

To analyze the properties of the Rayleigh wave, we separate $G^{\rm H}_{ij}$ into two parts:
\begin{align*}
G^{\rm H}_{ij}=R^{\rm H}_{ij}+O^{\rm H}_{ij},
\end{align*}
where
\begin{align*}
R^{\rm H}_{ij}=&H\left(t-t_{_{\rm P}}\right)\big[u_{ij,5}^{\rm P}U_{2}^{\rm P}\left(a^{\rm P}_{3}\right)+u_{ij,5}^{\rm P}U_{2}^{\rm P}\left(a^{\rm P}_{3}\right)+v_{ij,5}^{\rm P}V_{2}^{\rm P}\left(a^{\rm P}_{3}\right)+v_{ij,5}^{\rm P}V_{2}^{\rm P}\left(a^{\rm P}_{3}\right)\big] \nonumber \\
&+H\left(t-t_{_{\rm S}}\right)\big[u_{ij,5}^{\rm S}U_{2}^{\rm S}\left(a^{\rm S}_{3}\right)+u_{ij,5}^{\rm S}U_{2}^{\rm S}\left(a^{\rm S}_{3}\right)+v_{ij,5}^{\rm S}V_{2}^{\rm S}\left(a^{\rm S}_{3}\right)+v_{ij,5}^{\rm S}V_{2}^{\rm S}\left(a^{\rm S}_{3}\right)\big],\\
O^{\rm H}_{ij}=&G^{\rm H}_{ij}-R^{\rm H}_{ij}.
\end{align*}

We define $R^{\rm H}_{ij}$ as the Rayleigh term, because it contains all the terms which involve the actual zero of the Rayleigh function. Correspondingly, $O^{\rm H}_{ij}$ is referred to as the ``Other'' term. For other term, in Figs. 3 (b), 4 (b), 5 (b) and 6 (b), the displacements always decay smoothly after the arrival time of S wave. For Rayleigh term, however, in Figs. 3 (b) and 5 (b), there is an obvious disturbance at the arrival time of the Rayleigh wave. In Figs. 4 (b) and 6 (b), when the propotion of the depth of the source to the distance between the receiver and the source is small, the Rayleigh wave becomes the dominant phase.
The numerical analysis verifies that the Rayleigh wave is generated by $R_{ij}^{\rm H}$ and we shall report on additional features of the Rayleigh wave in subsequent articles on this issue. 
\begin{figure}
\centering
\subfigure{
\begin{minipage}{.48\textwidth}
\centering
\includegraphics[scale=0.55]{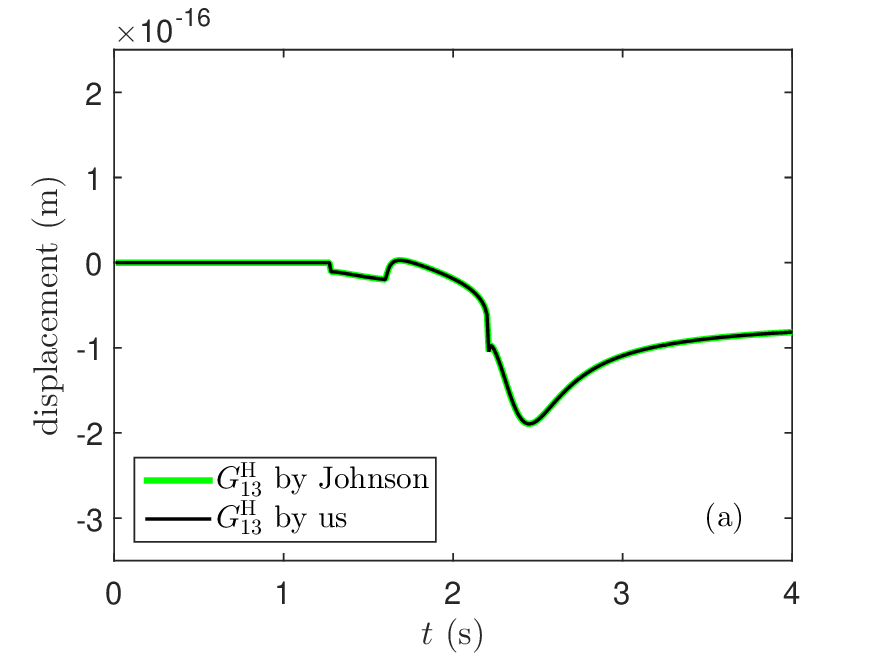}
\end{minipage}
}
\subfigure{
\begin{minipage}{.48\textwidth}
\centering
\includegraphics[scale=0.55]{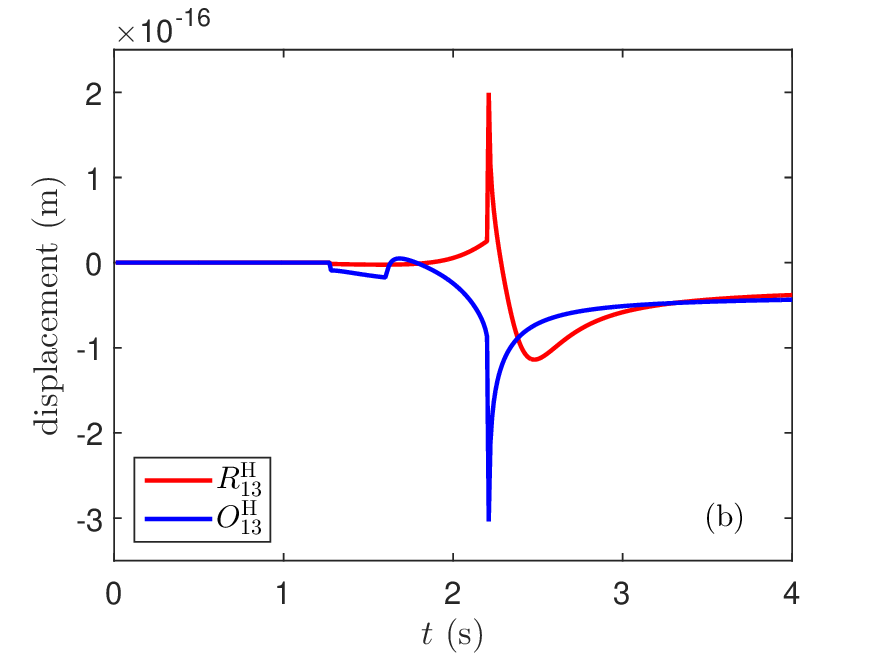}
\end{minipage}
}
\caption{(a) Comparison of $G^{\rm H}_{13}\left(10,0,0,t;0,0,2,0\right)$ between Johnson's ({\color{blue}1974}) solution and our solution; (b) $R^{\rm H}_{13}\left(10,0,0,t;0,0,2,0\right)$ and $O^{\rm H}_{13}\left(10,0,0,t;0,0,2,0\right)$.}
\end{figure}
\begin{figure}
\centering
\subfigure{
\begin{minipage}{.48\textwidth}
\centering
\includegraphics[scale=0.55]{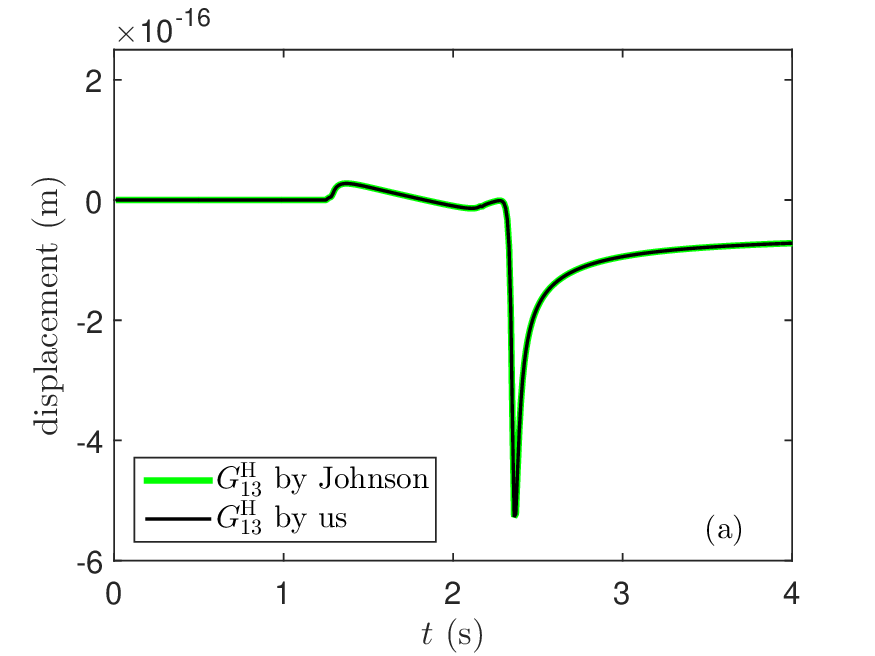}
\end{minipage}
}
\subfigure{
\begin{minipage}{.48\textwidth}
\centering
\includegraphics[scale=0.55]{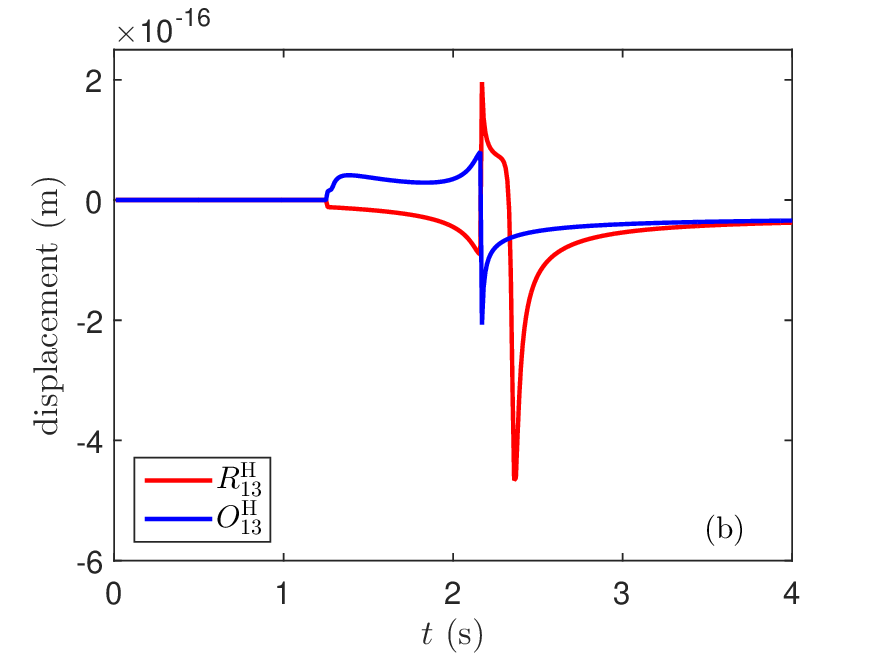}
\end{minipage}
}
\caption{(a) Comparison of $G^{\rm H}_{13}\left(10,0,0,t;0,0,0.2,0\right)$ between Johnson's ({\color{blue}1974}) solution and our solution; (b) $R^{\rm H}_{13}\left(10,0,0,t;0,0,0.2,0\right)$ and $O^{\rm H}_{13}\left(10,0,0,t;0,0,0.2,0\right)$.}
\end{figure}
\begin{figure}
\centering
\subfigure{
\begin{minipage}{.48\textwidth}
\centering
\includegraphics[scale=0.55]{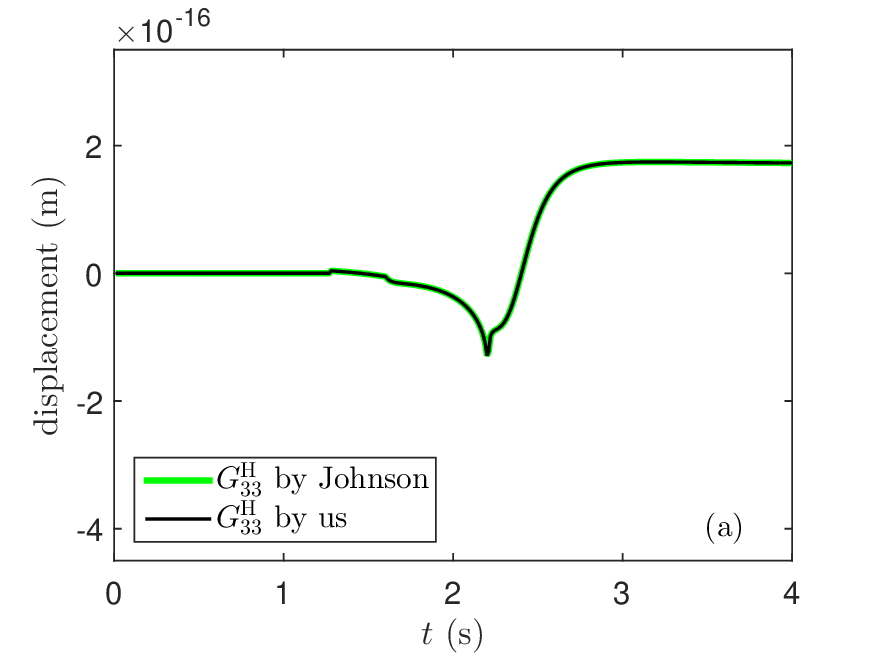}
\end{minipage}
}
\subfigure{
\begin{minipage}{.48\textwidth}
\centering
\includegraphics[scale=0.55]{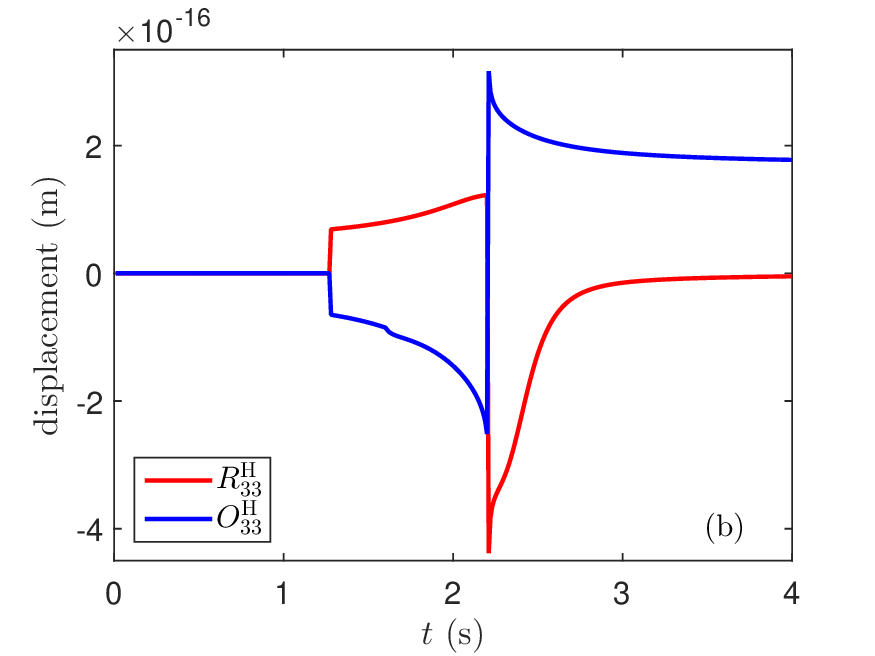}
\end{minipage}
}
\caption{(a) Comparison of $G^{\rm H}_{33}\left(10,0,0,t;0,0,2,0\right)$ between Johnson's ({\color{blue}1974}) solution and our solution; (b) $R^{\rm H}_{33}\left(10,0,0,t;0,0,2,0\right)$ and $O^{\rm H}_{33}\left(10,0,0,t;0,0,2,0\right)$.}
\end{figure}
\begin{figure}
\centering
\subfigure{
\begin{minipage}{.48\textwidth}
\centering
\includegraphics[scale=0.55]{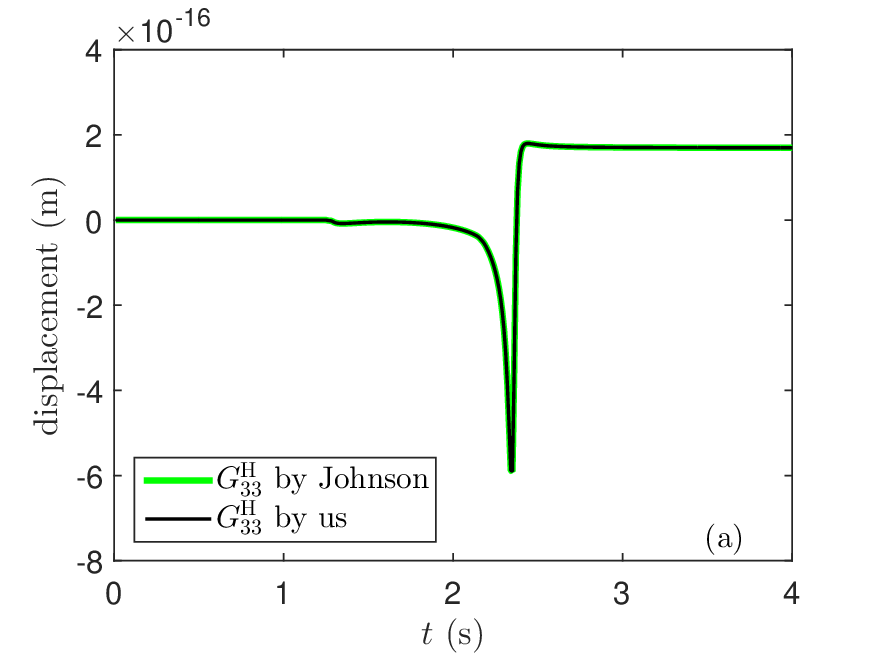}
\end{minipage}
}
\subfigure{
\begin{minipage}{.48\textwidth}
\centering
\includegraphics[scale=0.55]{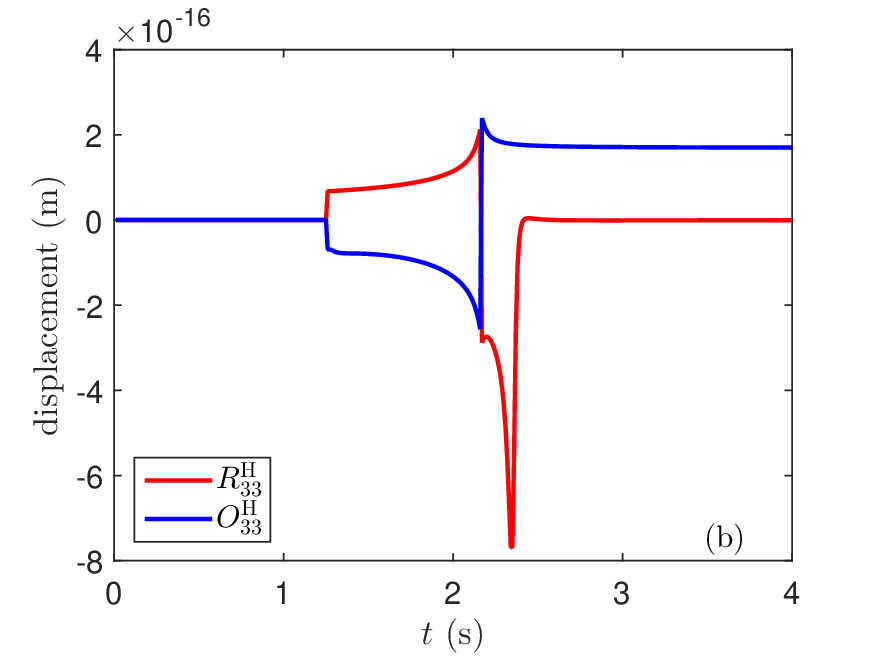}
\end{minipage}
}
\caption{(a) Comparison of $G^{\rm H}_{33}\left(10,0,0,t;0,0,0.2,0\right)$ between Johnson's ({\color{blue}1974}) solution and our solution; (b) $R^{\rm H}_{33}\left(10,0,0,t;0,0,0.2,0\right)$ and $O^{\rm H}_{33}\left(10,0,0,t;0,0,0.2,0\right)$.}
\end{figure}
\section{Conclusions}
In this article, we obtain the exact closed-form Green's function for Lamb's problem, which are the response on the free surface excited by a buried point force in a homogenuous elastic half-space. Our final expressions consist of the algebriac expressions and three kinds of complete elliptic functions, the form of which is similar to those in Richards ({\color{blue}1979}) and Kausel ({\color{blue}2012}), in which only a concentrated force located on the surface is considered. Our results are essentially equivalent to those of Johnson's ({\color{blue}1974}), for we manage to change his definite integrations into the simplest algebraic expressions elegantly. From a mathematical point of view, we give the classical Lamb's problem a successful and thorough answer.     
When we focus on the properties of a particular kind of waves such as the Rayleigh wave, our results also have a great advantage over Johnson's \shortcite{johnson}. In the forthcoming companion paper, we will indicate the outstanding features of the Rayleigh wave and the diffracted S-P wave from a buried point source and show the radiation pattern of body waves in the far field in the homogenuous half space based on the results obtained. The spatial derivatives of Green's function will also be calculated in the following papers. When the source and the receiver are both beneath the free surface, our method will still be valid with more complicated calculation. We will also fucus on the situation to distill more information about the Green's function in the half space.
\begin{acknowledgments}
This work was supported by the National Natural Science Foundation of China grant 41674050. We thank two anonymous reviewers for their useful suggestions which are crucial to the improvement of our manuscript.
\end{acknowledgments}
\balance
 
\nobalance
\appendix
\section{Expansion into partial fractions}
The Rayleigh function $R(B)$ can be factored as $R(B)=A\left(B^{2}-a_{1}^{2}\right)\left(B^{2}-a_{2}^{2}\right)\left(B^{2}+a_{3}^{2}\right)$, where $a_{1}$, $a_{2}$ and $a_{3}$ are all real constants and $A$ is s real constant.
For an arbitrary $n$-order ($n>6$) polynomial $P_{n}(B)$, the equality of the division algorithm is held as
\begin{align*}
P_{n}(B)=S_{n-6}(B)R(B)+Q_{5}(B),
\end{align*}
where $S_{n-6}(B)$ is a polynomial up to order $n-6$
in $B$ (or less) and $Q_{5}(B)$ is a polynomial up to order $5$
in $B$ (or less). So,
\begin{align*}
\frac{P_{n}(B)}{R(B)}=S_{n-6}(B)+\frac{Q_{5}(B)}{R(B)}.
\end{align*}
Consider the ratio of polynomials $\frac{Q_{5}(B)}{R(B)}$,
\begin{align}
\frac{Q_{5}(B)}{R(B)}&=\frac{\sum_{i=0}^{5}b_{i}B^{i}}{A\left(B^{2}-a_{1}^{2}\right)\left(B^{2}-a_{2}^{2}\right)\left(B^{2}+a_{3}^{2}\right)} \nonumber\\
&=\frac{1}{A}\frac{b_{0}+b_{2}B^{2}+b_{4}B^{4}}{\left(B^{2}-a_{1}^{2}\right)\left(B^{2}-a_{2}^{2}\right)\left(B^{2}+a_{3}^{2}\right)}+\frac{1}{A}\frac{\left(b_{1}+b_{3}B^{2}+b_{5}B^{4}\right)B}{\left(B^{2}-a_{1}^{2}\right)\left(B^{2}-a_{2}^{2}\right)\left(B^{2}+a_{3}^{2}\right)},
\end{align}
where $b_{i}$ $(i=0,...,5)$ are real constants. \\
The first part can be expanded as
\begin{align*}
&\frac{b_{0}+b_{2}B^{2}+b_{4}B^{4}}{\left(B^{2}-a_{1}^{2}\right)\left(B^{2}-a_{2}^{2})(B^{2}+a_{3}^{2}\right)}=\frac{c_{1}}{B^{2}-a_{1}^2}+\frac{c_{2}}{B^{2}-a_{2}^2}+\frac{c_{3}}{B^{2}+a_{3}^2} \nonumber\\
&=\frac{c_{1}}{2a_{1}}\left(\frac{1}{B-a_{1}}-\frac{1}{B+a_{1}}\right)+\frac{c_{2}}{2a_{2}}\left(\frac{1}{B-a_{2}}-\frac{1}{B+a_{2}}\right)+\frac{c_{3}}{B^{2}+a_{3}^2},
\end{align*}
where
\begin{align*}
&c_{1}=\frac{b_{0}+b_{2}a_{1}^{2}+b_{4}a_{1}^{4}}{\left(a_{1}^{2}-a_{2}^{2}\right)\left(a_{1}^{2}+a_{3}^{2}\right)},&&
c_{2}=\frac{b_{0}+b_{2}a_{2}^{2}+b_{4}a_{2}^{4}}{\left(a_{2}^{2}-a_{1}^{2}\right)\left(a_{2}^{2}+a_{3}^{2}\right)},&&
c_{3}=\frac{b_{0}-b_{2}a_{3}^{2}+b_{4}a_{3}^{4}}{\left(a_{3}^{2}-a_{1}^{2}\right)\left(a_{3}^{2}-a_{2}^{2}\right)}.
\end{align*}
Similarly, the other part can be expanded as
\begin{align*}
&\frac{\left(b_{1}+b_{3}B^{2}+b_{5}B^{4}\right)B}{\left(B^{2}-a_{1}^{2}\right)\left(B^{2}-a_{2}^{2}\right)\left(B^{2}+a_{3}^{2}\right)}=\frac{d_{1}B}{B^{2}-a_{1}^2}+\frac{d_{2}B}{B^{2}-a_{2}^2}+\frac{d_{3}B}{B^{2}+a_{3}^2} \nonumber\\
&=\frac{d_{1}}{2}\left(\frac{1}{B-a_{1}}+\frac{1}{B+a_{1}}\right)+\frac{d_{2}}{2}\left(\frac{1}{B-a_{2}}+\frac{1}{B+a_{2}}\right)+\frac{d_{3}B}{B^{2}+a_{3}^2},
\end{align*}
where
\begin{align*}
&d_{1}=\frac{b_{1}+b_{3}a_{1}^{2}+b_{5}a_{1}^{4}}{\left(a_{1}^{2}-a_{2}^{2}\right)\left(a_{1}^{2}+a_{3}^{2}\right)},&&
d_{2}=\frac{b_{1}+b_{3}a_{2}^{2}+b_{5}a_{2}^{4}}{\left(a_{2}^{2}-a_{1}^{2}\right)\left(a_{2}^{2}+a_{3}^{2}\right)},&&
d_{3}=\frac{b_{1}-b_{3}a_{3}^{2}+b_{5}a_{3}^{4}}{\left(a_{3}^{2}-a_{1}^{2}\right)\left(a_{3}^{2}-a_{2}^{2}\right)}.
\end{align*}
Eventually, the ratio of polynomials $\frac{P_{n}(B)}{R(B)}$ can be expansed as
\begin{align}
\frac{P_{n}(B)}{R(B)}=S_{n-6}(B)+\frac{u_{1}}{B-a_{1}}+\frac{u_{2}}{B+a_{1}}+\frac{u_{3}}{B-a_{2}}+\frac{u_{4}}{B+a_{2}}+\frac{u_{5}}{B^{2}+a_{3}^{2}}+\frac{u_{6}B}{B^{2}+a_{3}^{2}},
\end{align}
where
\begin{align*}
&u_{1}=\frac{1}{2A}(d_{1}+\frac{c_{1}}{a_{1}}) ,&& u_{2}=\frac{1}{2A}(d_{1}-\frac{c_{1}}{a_{1}}) ,&& u_{3}=\frac{1}{2A}(d_{2}+\frac{c_{2}}{a_{2}}) ,\\
&u_{4}=\frac{1}{2A}(d_{2}-\frac{c_{2}}{a_{2}}) ,&& u_{5}=\frac{c_{3}}{A} ,&& u_{6}=\frac{d_{3}}{A}.
\end{align*}
\section{Calculation of $U_{2}^{\rm P}$ and $U_{\rm 3}^{\rm P}$}
We introduce four parameters $A_{1}$, $A_{2}$, $\theta_{1}$ and $\theta_{2}$, that satisty
\begin{align*}
A_{1}e^{i\theta_{1}}=a+{\rm i}T_{\rm P}\cos{\theta}+\sqrt{T_{\rm P}^{2}-1}\sin{\theta},\\
A_{2}e^{i\theta_{2}}=a+{\rm i}T_{\rm P}\cos{\theta}-\sqrt{T_{\rm P}^{2}-1}\sin{\theta}.
\end{align*}
Utilizing the variable substitution $t=\tan{\frac{x}{2}}$ and the elementary integral formula
\begin{align*}
\int_{0}^{1}\left(mt^{2}+n\right)^{-1}\mathrm{d}t=\frac{1}{2{\rm i}\sqrt{mn}}\ln{\frac{\sqrt{n}+{\rm i}\sqrt{m}}{\sqrt{n}-{\rm i}\sqrt{m}}},
\end{align*}
we obtain
\begin{align*}
U_{2}^{\rm P}\left(a\right)&=\int_{0}^{\pi/2} \frac{1}{2a}\left(\frac{1}{a+{\rm i}B}+\frac{1}{a-{\rm i}B}\right)\mathrm{d}x\nonumber\\
&=\frac{1}{2a}\int_{0}^{\pi/2}\left(\frac{1}{a+{\rm i}T_{\rm P}\cos{\theta}-\sqrt{T_{\rm P}^{2}-1}\sin{\theta}\cos{x}}+\frac{1}{a-{\rm i}T_{\rm P}\cos{\theta}+\sqrt{T_{\rm P}^{2}-1}\sin{\theta}\cos{x}}\right)\mathrm{d}x\nonumber\\
&=\frac{1}{a}\int_{0}^{1}{\left[A_{1}e^{{\rm i}\theta_{1}} t^{2}+A_{2}e^{{\rm i}\theta_{2}}\right]^{-1}+\left[A_{2}e^{-{\rm i}\theta_{2}} t^{2}+A_{1}e^{-{\rm i}\theta_{1}}\right]^{-1}}\mathrm{d}t\nonumber\\
&=\frac{1}{2a{\rm i}\sqrt{A_{1}A_{2}}}e^{-{\rm i}\frac{\theta_{1}+\theta_{2}}{2}}\ln{\frac{\sqrt{A_{2}}+\sqrt{A_{1}}{\rm i}e^{{\rm i}\frac{\theta_{1}-\theta_{2}}{2}}}{\sqrt{A_{2}}-\sqrt{A_{1}}{\rm i}e^{{\rm i}\frac{\theta_{1}-\theta_{2}}{2}}}}+\frac{1}{2a{\rm i}\sqrt{A_{1}A_{2}}}e^{{\rm i}\frac{\theta_{1}+\theta_{2}}{2}}\ln{\frac{\sqrt{A_{1}}+\sqrt{A_{2}}{\rm i}e^{{\rm i}\frac{\theta_{1}-\theta_{2}}{2}}}{\sqrt{A_{1}}-\sqrt{A_{2}}{\rm i}e^{{\rm i}\frac{\theta_{1}-\theta_{2}}{2}}}} \nonumber\\
&=\frac{1}{2a{\rm i}\sqrt{A_{1}A_{2}}}e^{-{\rm i}\frac{\theta_{1}+\theta_{2}}{2}}M+\frac{1}{2a{\rm i}\sqrt{A_{1}A_{2}}}e^{{\rm i}\frac{\theta_{1}+\theta_{2}}{2}}N,
\end{align*}
where
\begin{align*}
M=\ln{\frac{\sqrt{A_{2}}+\sqrt{A_{1}}{\rm i}e^{{\rm i}\frac{\theta_{1}-\theta_{2}}{2}}}{\sqrt{A_{2}}-\sqrt{A_{1}}{\rm i}e^{{\rm i}\frac{\theta_{1}-\theta_{2}}{2}}}},\\
N=\ln{\frac{\sqrt{A_{1}}+\sqrt{A_{2}}{\rm i}e^{{\rm i}\frac{\theta_{1}-\theta_{2}}{2}}}{\sqrt{A_{1}}-\sqrt{A_{2}}{\rm i}e^{{\rm i}\frac{\theta_{1}-\theta_{2}}{2}}}}.
\end{align*}
Noticing that
$\Re\{M\}=\Re\{N\}$ and $\Im\{M\}+\Im\{N\}=\pi$, the final results can be simplyfied as
\begin{align*}
\Re{\int_{0}^{\pi/2} \frac{\mathrm{d}x}{B^{2}+a^{2}}}=\frac{\pi}{2}\frac{\cos{\frac{\theta_{1}+\theta_{2}}{2}}}{a\sqrt{A_{1}A_{2}}}.
\end{align*}
Similarly,
\begin{align*}
\Re{\int_{0}^{\pi/2} \frac{B \mathrm{d}x}{B^{2}+a^{2}}}=\frac{\pi}{2}\frac{\sin{\frac{\theta_{1}+\theta_{2}}{2}}}{\sqrt{A_{1}A_{2}}},
\end{align*}
where
$A_{1}A_{2}e^{{\rm i}(\theta_{1}+\theta_{2})}=a^{2}+\sin^{2}{\theta}-T_{\rm P}^{2}+2{\rm i}T_{\rm P}\cos{\theta}a$.
\label{lastpage}
\end{document}